\documentclass[reprint,aps,longbibliography,nofootinbib,floatfix,nobalancelastpage,superscriptaddress]{revtex4-2}
\usepackage{amsmath}
\usepackage{amsfonts}
\usepackage{amssymb}
\usepackage{mathtools}
\usepackage{graphicx}
\usepackage[dvipsnames]{xcolor}
\definecolor{myRed}{RGB}{214,39,40}

\usepackage[colorlinks=True,citecolor=RoyalPurple,linkcolor=Red,urlcolor=Violet]{hyperref}
\usepackage{braket}
\usepackage{hypcap}
\usepackage{yfonts}
\usepackage{bm}
\usepackage{ulem}
\usepackage{dsfont}
\newcommand\eq[1]{\begin{align}#1\end{align}}
\newcommand\ski{{S_{K,\infty}}}
\newcommand\skiq{{S^{(q)}_{K,\infty}}}
\newcommand\nh{N_\mathcal{H}}
\newcommand\ske{S_{K,\ket{E}}}
\newcommand\ze{z_{\ket{E}}}

\newcommand\new[1]{{#1}}

\newcommand{\pd}{\phantom\dagger}

\newcommand{\comment}[1]{}

\newcommand{\xibd}{\xi_{\mathrm{d}}}
\newcommand{\zb}{\overline{\xi}}

\begin{document}

\title{Krylov-space anatomy and spread complexity of a disordered quantum spin chain}
\author{Bikram Pain}
\email{bikram.pain@icts.res.in}
\affiliation{International Centre for Theoretical Sciences, Tata Institute of Fundamental Research, Bengaluru 560089, India}
\author{David E. Logan}
\email{david.logan@chem.ox.ac.uk}
\affiliation{University of Oxford, Physical and Theoretical Chemistry, South Parks Road, Oxford OX13QZ, United Kingdom }
\author{Sthitadhi Roy}
\email{sthitadhi.roy@icts.res.in}
\affiliation{International Centre for Theoretical Sciences, Tata Institute of Fundamental Research, Bengaluru 560089, India}
% \date{\today}

\begin{abstract}
We investigate the anatomy and complexity of quantum states in Krylov space, in the ergodic and many-body localised (MBL) phases of a disordered, interacting spin chain. The Krylov basis generated by the Hamiltonian from an initial state provides a 
representation in which the spread of the time-evolving state constitutes a basis-optimised measure of complexity. We show that the long-time Krylov spread complexity sharply distinguishes the two phases. In the ergodic \new{regime}, the infinite-time complexity scales linearly with the Fock-space dimension, indicating that the state spreads over a finite fraction of the Krylov chain. By contrast, it grows sublinearly in the MBL \new{regime}, implying that the long-time state occupies only a vanishing fraction of the chain. Further, the profile of the infinite-time state along the Krylov chain exhibits a stretched-exponential decay in the MBL 
\new{regime}. This behaviour reflects a broad distribution of decay lengthscales, associated with different eigenstates contributing to the long-time state. Consistently, a large-deviation analysis of the statistics of eigenstate spread complexities shows that while the ergodic \new{regime} receives contributions from almost all eigenstates, the complexity in the MBL \new{regime} is dominated by a vanishing fraction of eigenstates, which have anomalously large complexity relative to the typical ones.
\end{abstract}

\maketitle

\tableofcontents

\section{Introduction \label{sec:intro}}

The many-body localisation transition~\cite{oganesyan2007localisation,pal2010many} in disordered quantum many-body systems can be viewed as an eigenstate transition: across a critical disorder strength, the nature of eigenstates at arbitrary finite energy densities changes qualitatively between the ergodic and many-body localised 
(MBL) phases (see Refs.~\cite{nandkishore2015many,abanin2017recent,abanin2019colloquium,alet2018many, tikhonov2021anderson, roy2024fock, SierantReview2025} for reviews and further references). One of the most prominent real-space signatures of this transition is encoded in the eigenstate expectation values of local observables. In the ergodic phase, these satisfy the eigenstate thermalisation hypothesis~\cite{deutsch1991quantum,srednicki1994chaos,rigol2008thermalisation,dalessio2016from,deutsch2018eigenstate}, whereas in the MBL phase they violate it.

From a purely eigenstate perspective, the distinction between the ergodic and MBL phases is also reflected in bipartite entanglement entropy. Eigenstates in the ergodic phase typically exhibit volume-law entanglement, while \new{MBL eigenstates} obey an area law~\cite{bauer2013area,serbyn2013local,huse2014phenomenology,nandkishore2015many,abanin2019colloquium}. The latter is intimately connected to the existence of an extensive number of (quasi)local integrals of motion~\cite{serbyn2013local,huse2014phenomenology,ros2015integrals}. Consequently, MBL eigenstates can be related to trivial product states by finite-depth local unitary circuits, whereas no such local circuit exists for ergodic eigenstates~\cite{bauer2013area}. This perspective suggests that eigenstates in the two phases possess fundamentally different complexities, at least in the abstract sense of circuit complexity~\cite{ccdef,bernstein1997quantum,brandao2021models}

A complementary and equally insightful approach to the many-body localisation problem 
exploits the exact mapping of any disordered interacting many-body Hamiltonian to a tight-binding Hamiltonian, describing a fictitious single particle hopping on the 
associated correlated Fock-space graph of the system; where sites correspond to many-body basis states and links correspond to matrix elements of the Hamiltonian~\cite{logan1990quantum,altshuler1997quasiparticle,logan2019many,roy2020fock,tikhonov2021eigenstate,roy2021fockspace,detomasi2020rare,sutradhar2022scaling,ghosh2024scaling} (see Refs.~\cite{tikhonov2021anderson,roy2024fock} for reviews of this approach). The structure of many-body eigenstates on this graph is qualitatively different in the ergodic and MBL phases, as reflected e.g.\ in their inverse participation ratios 
(IPRs). Specifically, MBL eigenstates display multifractal statistics, manifest in anomalous scaling of the IPRs with the Fock-space dimension $\nh$, while ergodic eigenstates are fully extended over the graph~\cite{deluca2013ergodicity, luitz2015many, mace2019multifractal,tikhonov2021eigenstate,roy2021fockspace,roy2025spectral}. Physically, this implies that MBL eigenstates occupy a number of Fock-space sites which, although exponentially large in system size, constitutes a vanishing fraction of the total Fock-space dimension, while ergodic eigenstates by contrast occupy a finite fraction of the graph.

A dynamical manifestation of this distinction arises when the system is initialised in a many-body basis state corresponding to a single site of the Fock-space graph. Under time evolution, the state spreads over the graph. In the ergodic phase, the long-time state occupies a finite fraction of the graph, whereas in the MBL \new{regime} it remains confined to a vanishing fraction even at arbitrarily late times~\cite{creed2023probability,pain2023connection}. This suggests the existence of a notion of complexity associated with the spread of the state on the Fock-space graph, which should behave qualitatively differently in the two phases.

An intuitive measure of this complexity is the size of the wavefunction on the graph.
But such a measure is basis dependent, so this naturally raises the question of whether there exists a basis in which the spread of the state -- and hence the associated complexity -- is minimised. Importantly, it was shown that the Krylov basis generated by the Hamiltonian from the initial state minimises the complexity of the time-evolving state~\cite{bala2022quantum}. Consequently, the spread complexity of the state in Krylov space (hereafter referred to as the Krylov spread complexity) provides a bona fide, basis-optimised measure of complexity. 

This in turn motivates the central question of this work: what is the anatomy of states in Krylov space in the ergodic and MBL phases, and how does the Krylov spread complexity distinguish between them? We also show that the Krylov-space anatomy of states provides insights into the MBL \new{regime} that are arguably  more difficult to 
obtain from the Fock-space perspective. This is because the Hamiltonian in Krylov space takes the form of a one-dimensional (correlated) disordered tight-binding chain of length $\nh$, where the ordering of basis states and the notion of distance are much simpler than on the high-dimensional Fock-space graphs.

The present work should be contrasted with the much more widely studied operator Krylov complexity~\cite{parker2019universal,dymarsky2020quantum,dymarsky2021krylov,rabinovici2021operator,rabinovici2022krylov,trigueros2022Krylov,alishahiha2023universal,Menzler2024Krylov,avdoshkin2024krylov,campo2024shortcuts, suchsland2025krylov,chen2025dissecting,chapman2025krylov,NANDY2025}, where the operator growth of an initially real-space-local operator is mapped onto the spread of the operator in the Krylov space of the operator Hilbert space.  In fact, the operator Krylov complexity is not necessarily a sensitive probe of the MBL \new{regime, because here too}  the operator may delocalise in the Krylov space, as shown e.g.\ for the phenomenological $\ell$-bit model~\cite{trigueros2022Krylov} (neither in fact is it a sensitive probe of integrability~\cite{bhattacharjee2022krylov}).

On the other hand, as shown in this work, the Krylov-space anatomy, as well as the Krylov spread complexity of an initial state~\cite{bhattacharjee2025krylovRMT,ganguli2024state,cohen2024complexity,CamargoJHEP2024,baggioli2025krylov,alishahiha2025Krylov,Peacock2026AndersonKrylov,bala2025quantum,huh2025krylov,camargo2025higher,scialchi2024integ} evolved for arbitrarily long (infinite) times, shows qualitative differences in the ergodic and MBL \new{regimes}. In fact, the Krylov spread complexity has already proven successful in capturing aspects of ergodicity breaking in 
both single-particle and many-body settings under specific initial conditions~\cite{bhattacharjee2025krylovRMT,ganguli2024state,cohen2024complexity,CamargoJHEP2024,baggioli2025krylov,alishahiha2025Krylov,Peacock2026AndersonKrylov}.
More specifically, we find that the infinite-time Krylov spread complexity in the ergodic \new{regime} scales linearly with the Fock-space dimension, $\propto \nh$, whereas in the MBL \new{regime} it grows sublinearly as $\nh^{\alpha}$ with $\alpha<1$. Coupled to the fact that the Krylov space is an ordered one-dimensional chain of length $\nh$, the above scaling presents a simple yet insightful picture: in the ergodic \new{regime} the infinite-time states extend over a finite fraction  of the entire Krylov chain, while in the MBL \new{regime}, they extend to only a vanishing fraction of the chain. 

In the MBL \new{regime} we find in fact that the profile of the infinite-time state along the length of the Krylov chain has a non-trivial decay described by a stretched exponential. We present a phenomenological theory which shows how such a profile can emerge, from the interplay of exponential decay of the state amplitude on the Krylov chain and a broad (but not itself heavy-tailed) distribution of associated decay length scales. Importantly, these distributions arise not only over disorder realisations, but also across eigenstates within a given realisation. This interplay becomes evident when the infinite-time Krylov spread complexity is expressed as a sum over eigenstates of the Krylov spread complexities associated with individual eigenstates.

The latter decomposition yields further insight into the nature of eigenstates in the two \new{regime}. In the ergodic \new{case}, eigenstate spread complexities are found to
have a narrow distribution over the eigenstates. By contrast, in the MBL \new{regime} the distribution is broad, and its tails become heavier with increasing system size. Using a large-deviation analysis, we show that in the ergodic \new{regime} the infinite-time Krylov spread complexity receives contributions from almost all eigenstates, whereas in the MBL \new{regime} it is dominated by only a vanishing fraction of eigenstates -- albeit exponentially many in system size --  lying in the tails of the distribution. This provides a  manifestation, on the Krylov chain, of rare resonances characteristic of the MBL \new{regime}~\cite{detomasi2020rare,morningstar2021avalanches,garratt2021resonances,garratt2022resonant,crowley2022constructive,ha2023manybody,colbois2024statistics,colbois2024interaction,biroli2024largedeviation,colbois2025cat,padhan2025long,miranda2025large}.

The paper is organised as follows. Sec.~\ref{sec:defn}  reviews briefly the definition of the Krylov spread complexity and the construction of the Krylov basis. 
In Sec.~\ref{sec:model-kc}, we specify the disordered, tilted-field Ising chain as our model of choice, and discuss its Krylov chain and its conceptual connections to the model's Fock-space graph. Detailed numerical results for the Krylov-space anatomy and the spread complexity, along with a phenomenological theory in the MBL \new{regime}, constitute Sec.~\ref{sec:ski-lam-n-res}. The large-deviation analysis of the statistics of the eigenstate spread complexities contributing to the infinite-time spread complexity is presented in Sec.~\ref{sec:stats-spread}. Some concluding remarks are given in Sec.~\ref{sec:conclusion}.

\section{Krylov spread complexity \label{sec:defn}}

We start by discussing the construction of the Krylov chain, and formally defining the associated spread complexity~\cite{bala2022quantum}. This enables us to discuss several important conceptual points, and to put on a concrete footing the motivation behind studying 
the anatomy of states on the Krylov chain.

 Consider a general many-body Hamiltonian of form
\eq{
H = H_0^{\pd} + H_1^{\pd}\,,
\label{eq:ham-general}
}
denoting by $\{\ket{I}\}$ the eigenstates of $H_0$, and by $\{\ket{E}\}$ the eigenstates of $H$ with eigenvalue $E$. The setting in which we will be interested
is initialising the state of system at $t=0$ in an eigenstate of $H_0$, $\ket{\psi_{t=0}} = \ket{I}$, and asking how {\it complex} the state becomes upon evolution over an arbitrarily long time under the full Hamiltonian $\ket{\psi_t} = e^{-i H t}\ket{\psi_0}$.  Denoting by $S_K(t)$ the measure of complexity for the state at time $t$, our specific focus will be the long-time limit 
$\ski \equiv \lim_{t\to\infty}S_{K}(t)$.
This can be expressed in terms of correlated eigenstate amplitudes on the Krylov chain, as we show below.

For a general state $\ket{\psi_t}$, given some orthonormal basis 
$\{\ket{{\cal V}_n}\}$, this can be quantified using a cost function
\eq{
C_{\cal V}^{\pd} = \sum_n \mu_n^{\pd} |\braket{\psi_t^{\pd}|{\cal V}_n^{\pd}}|^2\,,
\label{eq:cost}
}
where the $\mu_n$s are a positive, increasing sequence of real numbers. It is 
customary~\cite{bala2022quantum} to consider $\mu_n = n$, such that $C_{\cal V}$ can be interpreted as the size of the support of the state $\ket{\psi_t}$ assuming the basis 
$\{\ket{{\cal V}_n}\}$ to be ordered.
A natural measure of complexity is therefore to consider the basis which minimises the cost function. An important result in this context is that the basis which minimises Eq.~\ref{eq:cost} is the Krylov basis~\cite{bala2022quantum}, generated 
via the Hamiltonian $H$, with its first basis vector given by $\ket{\psi_0}$.
This is an ordered orthonormal basis, the basis states being denoted 
by $\ket{k_n}$ with $n=0,1,\cdots,\nh-1$, where $\nh$ is the Fock-space dimension of the system. These Krylov basis states are defined via the recursion relation~\cite{ViswanathRecursionbook}
\eq{
\ket{k_n^{\pd}} = \frac{1}{b_n}\big[H\ket{k_{n-1}^{\pd}} - a_{n-1}^{\pd}\ket{k_{n-1}^{\pd}} - b_{n-1}^{*}\ket{k_{n-2}^{\pd}}\big]\,,
\label{eq:krylov}
}
where $a_{n} = \braket{k_n|H|k_n}$ and
$b_n = \braket{k_{n}|H|k_{n-1}}$, and $\ket{k_0}=\ket{I}$ is the boundary condition for the recursion.

An immediate consequence of this construction is that the many-body Hamiltonian in the Krylov basis takes the tridiagonal form
\eq{
H = \smashoperator[r]{\sum_{n=0}^{\nh-1}}a_n^{\pd} \ket{k_n^{\pd}}\bra{k_n^{\pd}} + 
{\sum_{n=1}^{\nh -1}}[b_n^{\pd}\ket{k_{n-1}^{\pd}}\bra{k_n^{\pd}} + {\rm h.c.}]\,.
\label{eq:H-krylov-tridiag}
}
The Krylov basis is thus equivalent to a one-dimensional, nearest-neighbour, tight-binding chain of length $\nh$, described by the Hamiltonian Eq.~\ref{eq:H-krylov-tridiag}. While the fact that the length of the Krylov chain is $\nh$ is a
straightforward fallout,  its conceptual importance is that it provides us
with a naturally ordered one-dimensional basis for the many-body Hilbert space, 
which has a simple and unambiguous notion of distance. 

With the initial state $\ket{I}=\ket{k_0}$ localised at one end of the chain, 
the temporal evolution of the state on the Krylov chain is encoded in the 
probability amplitudes
\eq{
c_n^{\pd}(t) = \braket{k_n^{\pd}|\psi_t^{\pd}}=\sum_E e^{-iEt}\braket{k_n^{\pd}|E}\braket{E|k_0^{\pd}}\,.
\label{eq:cn(t)}
}
The spread complexity is then defined as~\cite{bala2022quantum} 
\eq{
S_K^{\pd}(t) = \sum_{n=0}^{\nh-1}n|c_n^{\pd}(t)|^2\,,
\label{eq:SK(t)}
}
and the one-dimensional nature of the Krylov chain leads to the interpretation of the spread complexity as simply the size of the support of the wavefunction on the chain.

We will be particularly interested in the spread complexity at infinite time, formally defined as
\eq{
S_{K,\infty}^{\pd} = \lim_{t\to\infty}\frac{1}{t}{\int_0^t} dt'~S_K^{\pd}(t')\,.
\label{eq:SKinfdef}
}
This can be conveniently expressed as
\eq{
S_{K,\infty}^{\pd} = \smashoperator[r]{\sum_{n=0}^{\nh-1}}n \Lambda_n^{\pd}\,,
\label{eq:SK-inf}
}
where (with $\ket{E}$ denoting an eigenstate of $H$)
\eq{
\Lambda_n^{\pd} = \sum_E\Lambda_{n,\ket{E}}^{\pd}~{\rm with}~\Lambda_{n,\ket{E}^{\pd}} = |\braket{k_n^{\pd}|E}\braket{E|k_0^{\pd}}|^2\,.
\label{eq:Lambda-n-E}
}
Equations~\ref{eq:SK-inf} and \ref{eq:Lambda-n-E}  relate the infinite-time spread complexity to eigenstate amplitudes on the Krylov chain. Note that conservation of probability trivially leads to $\Lambda_n$ satisfying the sum rule
\eq{
\sum_{n=0}^{\nh-1}\Lambda_n^{\pd} = 1\,.
\label{eq:Lambda-n-sum-rule}
}
As such, $\Lambda_n$ can interpreted as a probability distribution over the Krylov chain, for which $S_{K,\infty}$ is nothing but its first moment.
The profile of $\Lambda_n$ on the Krylov chain, and its scaling with
$\nh$, together with the statistical properties of the eigenstate spread complexity $S_{K,\infty}$, will be the central quantities of interest in 
this work.
\new{While $\ski$ is our focus in this work, it is of course worth 
noting that the temporal dynamics of $S_K(t)$ per se is extremely rich, and much is to be learned from it. Preliminary results presented in Appendix~\ref{app:dynamics} show that the dynamics of $S_K(t)$ is qualitatively different for weak and strong disorder. At weak disorder, 
 $S_K(t)$ is found to grow linearly with $t$ until reaching  saturation at the Heisenberg time $t_H\sim \nh$, characteristic of ergodic/chaotic phases; while at strong disorder on the other hand, the pre-saturation growth is anomalously slow. In either case, however, the long-time 
saturation value of $S_K(t)$ is indeed the $\ski$ obtained (as in Eqs.~\ref{eq:SKinfdef}-\ref{eq:Lambda-n-E}) by taking the infinite-time limit from the outset.
}

\section{Disordered spin-1/2 chain and its Krylov chain \label{sec:model-kc}}

\subsection{Model}
\label{sec:Model}

As a concrete setting for our studies, we consider the disordered tilted-field Ising (TFI) spin-1/2 chain, which has emerged as the standard archetype for a disordered system hosting a MBL phase at strong disorder~\cite{imbrie2016many,abanin2021distinguishing,SierantReview2025,roy2024fock}.
The model is described by the Hamiltonian
\eq{
H = \underbrace{\sum_{i=1}^{L-1}J_i^{\pd}\sigma^z_i\sigma^z_{i+1} + \sum_{i=1}^L h_i^{\pd}\sigma^z_i}_{H_0^{\pd}} + \underbrace{\Gamma\sum_{i=1}^L\sigma^x_i}_{H_1^{\pd}}\,,
\label{eq:ham-tfi}
}
where (taking $\Gamma \equiv 1$)
$J_i \in [0.8,1.2]$ and $h_i\in [-W,W]$ are independent random numbers. 
This model hosts an MBL phase for strong enough disorder, and an ergodic phase
at weaker disorder strengths. For the parameters above, within accessible
system sizes, the  critical disorder strength is
\new{estimated as} $W_c\simeq 3.7$~\cite{abanin2021distinguishing}.
\new{While the precise value of $W_c$ for certain specific models is under some debate and numerical works have suggested it to be 
significantly larger in the thermodynamic limit~\cite{suntajs2020quantum,morningstar2021avalanches,sels2022bath,peacock2024many}, the 
existence of the MBL phase in one spatial dimension is not under much debate~\cite{imbrie2016many,deroeck2025absencenormalheatconduction}.
Operationally, we will thus refer to the phenomenology observed for accessible system sizes at $W\gtrsim 4$ as the MBL regime, under the premise that the same phenomenology remains robust in a genuine MBL phase in the thermodynamic limit, albeit possibly at higher $W$.
}

Given that $H_0$ in Eq.~\ref{eq:ham-tfi} is composed solely of $\{\sigma^z_i\}$ operators, the basis $\{\ket{I}\}=\{\ket{\{\sigma^z_i\}}\}$ is simply the $\sigma^z$-product state basis. The $0^{\rm th}$ Krylov orbital, $\ket{k_0}$, is then simply a specific $\sigma^z$-product state, which we denote by $\ket{I_0}$.
\new{Starting from a $\sigma^z$-product state is the natural physical choice, because in the  model Eq.~\ref{eq:ham-tfi} the disorder couples to the $\sigma^z$ component of the spins and hence these states are precisely the MBL eigenstates in the limit of $W\to\infty$. At large but finite $W$ in the MBL regime, the MBL eigenstates are again connected to the $\sigma^z$-product states via finite-depth local unitary operators~\cite{bauer2013area}.}
In our numerical calculations, \new{for a given disorder realisation} we consider $\ket{I_0}$ to be the $\sigma^z$-product state whose energy is closest to the middle of the many-body 
spectrum.~\footnote{\new{This is just a matter of convenience:
results are unaffected if for a given disorder realisation one 
instead considers several initial product states near the middle of the spectrum (each of course with its own Krylov space), and accumulates statistics over them.}
}

With this choice, the entire set of Krylov-basis vectors $\{\ket{k_n}\}$ together with the parameters $\{a_n\}$ and $\{b_n\}$ describing the Krylov Hamiltonian,  can be computed explicitly. 
Some details of the statistical properties of the $\{a_n,b_n\}$ 
are relegated to Appendix~\ref{app:anbn}, but one essential  point should be mentioned here. Specifying an instance of the Hamiltonian in Eq.~\ref{eq:ham-tfi} requires only $\propto L$ independent random values of $J_i$ and $h_i$.
By contrast, specifying the Krylov Hamiltonian requires $\propto \nh$ values of $a_n$ and $b_n$ (Eq.~\ref{eq:H-krylov-tridiag}).
However, since the latter $\propto \nh$ coefficients are constructed from only 
$\propto L$ independent $J_i$ and $h_i$ values, strong correlations between the matrix elements of the Krylov Hamiltonian necessarily arise. As a result, the problem is very different from a conventional Anderson localisation problem~\cite{roy2020localisation,duthie2022anomalous,logan2025mf}, despite the Krylov chain being one dimensional.

\subsection{On the connection between Krylov and Fock-space bases}

Before delving into the anatomy of eigenstates on the Krylov chain, it
is important from a conceptual point of view to understand the connection between the Krylov basis and the more commonly employed Fock-space basis, and its associated Fock-space graph. 
 
The eigenstates of $H_0$ ($\sigma^z$-product states $\{\ket{I}\}$) form a natural choice of basis for the Fock space. Properties of the  Fock-space graph have been discussed in detail previously (see \cite{roy2024fock} and references therein); here we summarise  the features of it essential for the present work.

\begin{figure}
\includegraphics[width=\linewidth]{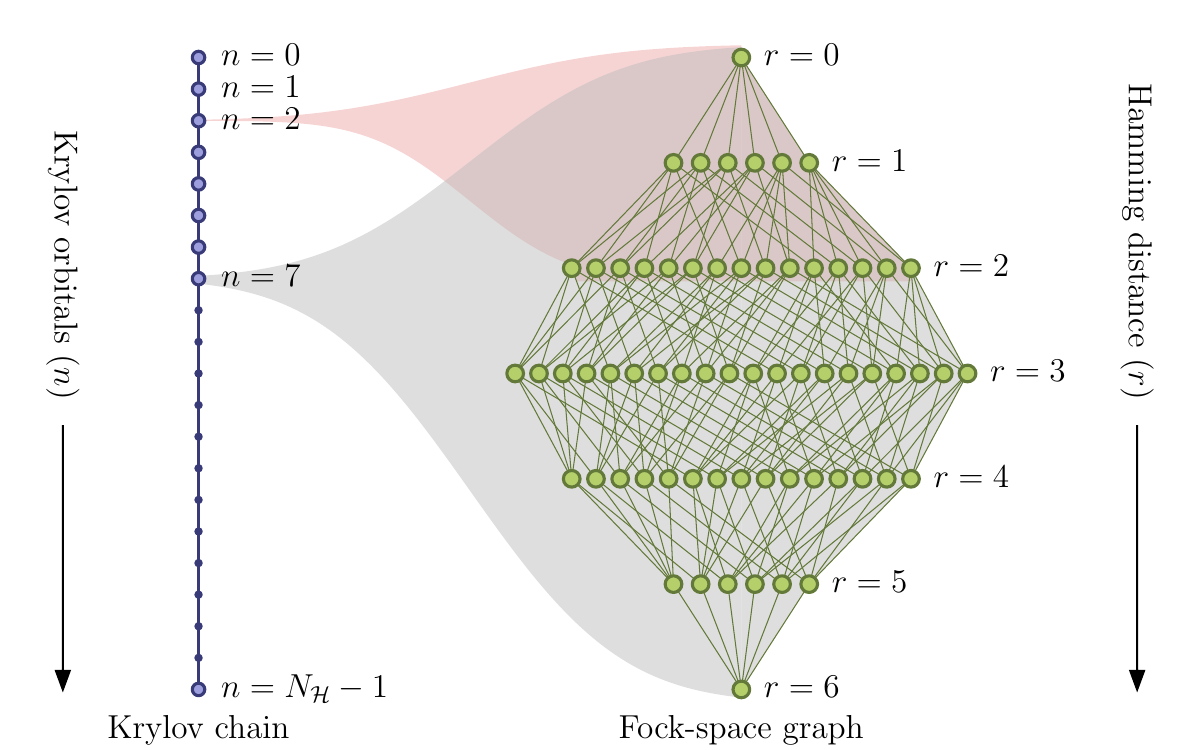}
\caption{
In the Krylov-space basis, the many-body Hamiltonian takes the form of a nearest-neighbour tight-binding chain (Eq.~\ref{eq:H-krylov-tridiag}), of length  
$\nh=2^L$, with the orbitals $\ket{k_n}$ labelled $n=0,1,\cdots,\nh-1$.
On the Fock-space graph (illustrated for $L=6$), the
Krylov $\ket{k_0}$ orbital is identified with the top apical site $\ket{I_{0}}$.
For $n\leq L$, $\ket{k_n}$ is supported on Fock-space sites which lie within Hamming distance $r\leq n$, as illustrated by the red-shaded connection between $\ket{k_2}$ and all Fock-space sites with $r\leq 2$. By contrast, for $L+1\leq n\leq \nh-1$, $\ket{k_n}$ is in general supported on the entire Fock-space graph, as indicated by the grey-shaded connections.}
\label{fig:schematic}
\end{figure}

The Fock-space graph for the TFI chain, illustrated as part of Fig.~\ref{fig:schematic}, is an $L$-dimensional hypercube with $\nh=2^L$ nodes, each such 
representing a $\sigma^z$ configuration. Each node has a connectivity of precisely $L$, which corresponds to flipping each of the $L$ spins on the chain under the action of $H_1$ ($=\Gamma\sum_{i=1}^L\sigma_{i}^{x}$). The Hamming distance between two $\sigma^z$ configurations -- defined as the number of sites where the two configurations differ from each other -- endows the Fock-space graph with a natural notion of distance. For the form of $H_1$ in Eq.~\ref{eq:ham-tfi}  it is also, conveniently, the shortest path between the two nodes. The above features 
imply that for a given $\ket{I_0}$, the remaining nodes can be arranged in a row-wise fashion such that all nodes on row $r$ are a Hamming distance of $r$ from $\ket{I_0}$.  Naturally there are $L$ such rows, with the $r^{\rm th}$ row containing $N_r=\binom{L}{r}$ nodes.  The form of $H_1$ in Eq.~\ref{eq:ham-tfi} also implies that any node on row $r$ is connected to nodes only on rows $r\pm 1$.

Given the structure of the Fock-space graph, a natural question is: what is the 
distribution of the Krylov-basis vectors, $\{\ket{k_n}\}$, on the Fock-space graph?
To make this quantitative, define the overlap of the Krylov-basis states and the Fock-space basis states as 
\eq{
w_{nI}^{\pd} = |\braket{k_n^{\pd}|I}|^2\,,
}
and the average weight of the $n^{\rm th}$ Krylov-basis state at  Hamming distance $r$ from $I_0$ on the Fock-space graph as 
\eq{
w_n^{\pd}(r)=\frac{1}{N_r}\sum_{I:r_{II_0}=r}w_{nI}^{\pd}= \frac{1}{\binom{L}{r}}\sum_{I:r_{II_0}=r}|\braket{k_n^{\pd}|I}|^2\,. 
\label{eq:wally}
}
Considering $\ket{k_0} = \ket{I_0}$, the construction of the Krylov chain in Eq.~\ref{eq:krylov}, together with the form of $H_1$, makes it clear that for $n\leq L$,
the Krylov $\ket{k_n}$ has support on Fock-space sites which lie at Hamming distances $r\leq n$ from $I_0$. However, for $n>L$, the orthonormality of $\ket{k_n}$ with $\ket{k_m}~\forall~m<n$ implies that the support of $\ket{k_n}$ folds back on the Fock-space graph, and thus in principle exists on the entire graph.  This is shown schematically in  Fig.~\ref{fig:schematic} for the case of $L=6$.
Indeed, since the Krylov chain has a length $\nh =2^L$, a fraction tending to unity of the sites thereon potentially have support  over the entire Fock space.

It is of course important to establish that the Krylov-basis states indeed span the 
Fock-space graph, whether in the ergodic or MBL regimes.
To this end, we define an effective size of the wave function corresponding to the Krylov orbital $\ket{k_n}$ on the Fock-space graph, via
\eq{\mathcal{R}_n^{\pd} = \sum_{r=0}^L r 
\binom{L}{r} w_n^{\pd}(r) = \sum_{r=0}^L r  \sum_{I: r_{II_0}=r}w_{nI}^{\pd}\,.
\label{eq:Rn}
}
Physically, $\mathcal{R}_{n}$ gives the mean Hamming distance on the Fock-space graph that is associated with the given Krylov $\ket{k_{n}}$, and as such is a bona fide measure of how the Krylov-basis state is spread over the Fock-space graph.
Since $\{\ket{k_n}\}$ forms a complete orthonormal basis,
such that $\sum_{n} w_{n}(r)=1$ (Eq.~\ref{eq:wally}), note that ${\cal R}_n$ 
satisfies a sum rule
\eq{
\sum_{n=0}^{\nh-1}{\cal R}_n^{\pd} 
=\sum_{r=0}^{L} r \binom{L}{r}
= \nh L/2\,.
\label{eq:Rn-sumrule}
}

\begin{figure}
\includegraphics[width=\linewidth]{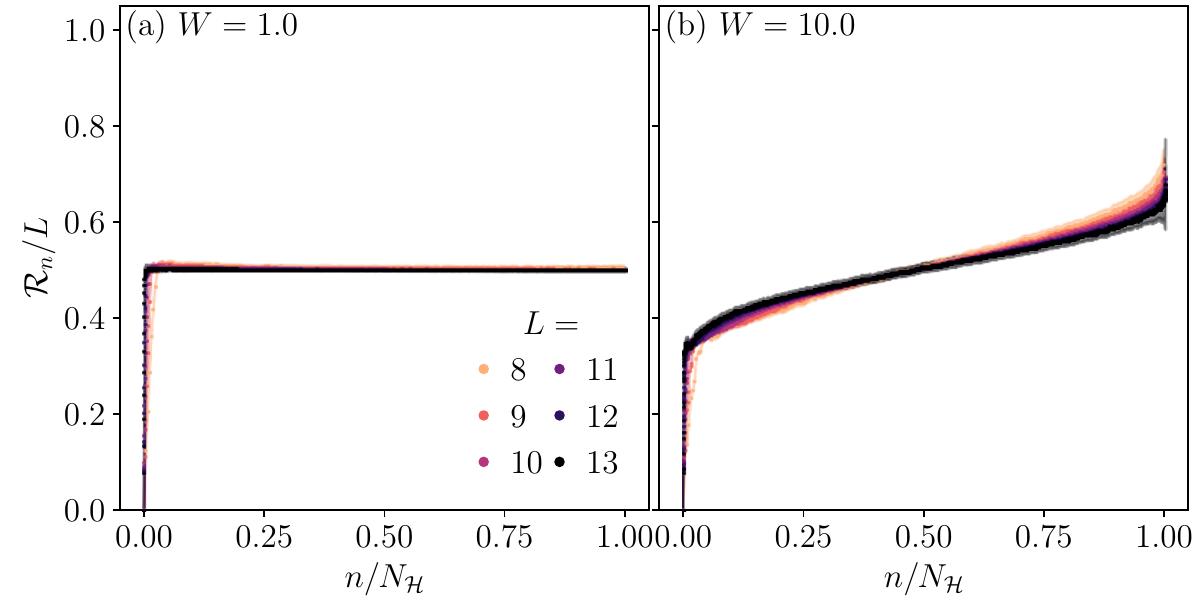}
\caption{
Numerical  results for the effective size of the wave function on the Fock-space graph corresponding to the $n^{\rm th}$ Krylov orbital, as quantified by ${\cal R}_n$ defined in Eq.~\ref{eq:Rn} (with system sizes $L$ indicated). Panels (a) and (b) with $W=1$ and $W=10$ exemplify, respectively, the ergodic and MBL regimes. The results conform to the scaling form in Eq.~\ref{eq:Rn-scaling}. The (thin) shaded region around the data points reflects the statistical errors over disorder realisations.}
\label{fig:Rn}
\end{figure}

Sufficiently deep in the ergodic phase, one expects $w_{nI}\sim \nh^{-1}$  and hence $w_n(r) \sim \nh^{-1}$ for all $I$ and $n>L$; which when used in Eq.~\ref{eq:Rn} yields ${\cal R}_n\simeq L/2$. That this is indeed the case is confirmed by numerical results shown in  
Fig.~\ref{fig:Rn}(a) for a disorder strength $W=1$ deep in the ergodic phase (and with system sizes $L$ as indicated).

The natural generalisation of this to encompass the MBL regime is the binomial form
\eq{
w_n^{\pd}(r) = p_n^r(1-p_n^{\pd})^{L-r}\,,
}
which from Eq.~\ref{eq:Rn}
yields ${\cal R}_n = p_nL$ (with the deep ergodic phase corresponding to $p_n=1/2$
$\forall n$).
The sum rule in Eq.~\ref{eq:Rn-sumrule} then implies $\sum_n p_n = \nh/2$.
Assuming $p_n = g(n/\nh^a)$ to be some function of $n/\nh^{a}$ (with 
exponent $a$), this sum rule  can be cast for $\nh\gg 1$ as
\eq{
\int_0^{\nh} dn~g\left(\frac{n}{\nh^a}\right) 
=
\nh^a\int_0^{\nh^{1-a}} dx~g(x)
= \frac{\nh}{2}\,,
}
which naturally requires $a=1$. This then implies that ${\cal R}_n$ satisfies the scaling form
\eq{
{\cal R}_n^{\pd} = L\times g\left(\frac{n}{\nh}\right)\,,
\label{eq:Rn-scaling}
}
such that $\mathcal{R}_{n}/L$ is a function solely of $n/\nh$.
While the scaling form is quite trivially satisfied in the ergodic regime (as 
above), numerical results in the MBL \new{regime}, shown in Fig.~\ref{fig:Rn}(b) 
for disorder strength $W=10$, confirm this to be case in general. The key upshot of this scaling form is that in both the ergodic and MBL regimes, the Krylov-basis states indeed have access to the entire Fock-space  graph~\footnote{This is not of course the case for the first $L$ Krylov basis states, but that is immaterial since these constitute an exponentially small (in $L$) fraction of the $2^{L}$ Krylov states.}.

\section{Scaling forms for spread complexity and eigenstate amplitudes
\label{sec:ski-lam-n-res}}

With the basic properties of the Krylov chain and the spread complexity defined, we turn now to the Krylov-space anatomy of the eigenstates.

\subsection{Distribution and scaling of $\ski$ \label{sec:ski-dist-mean}}

\begin{figure}
\includegraphics[width=\linewidth]{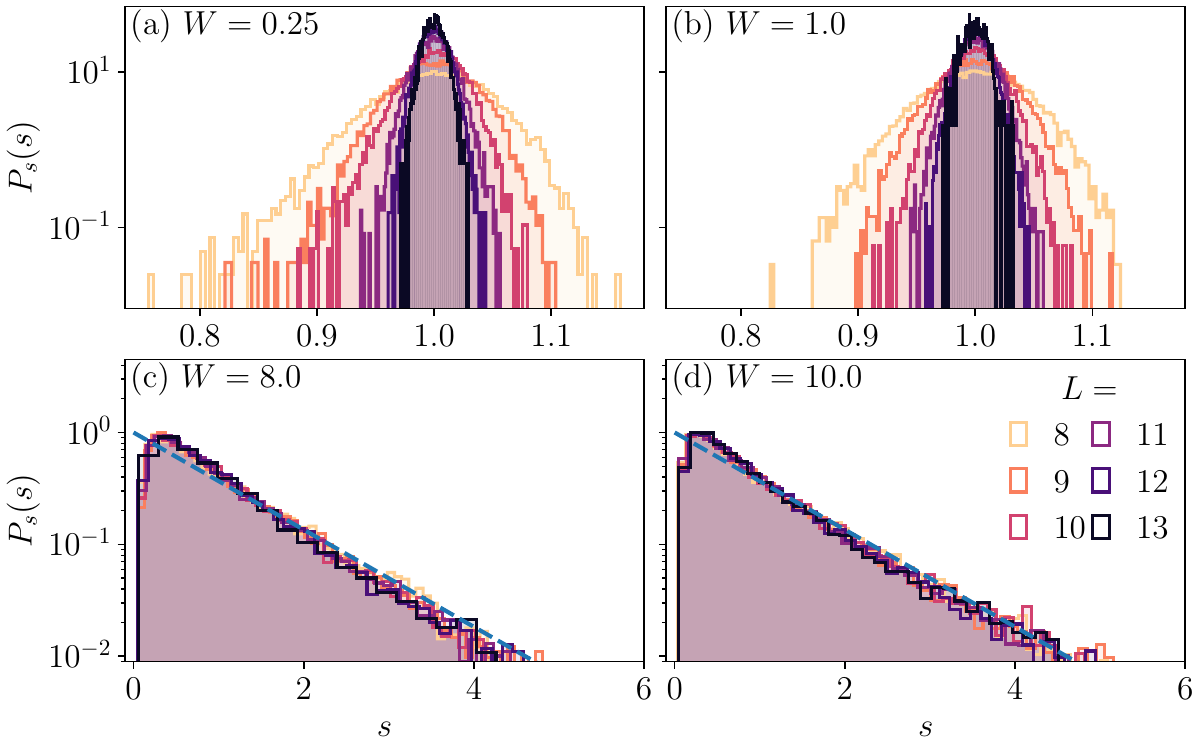}
\caption{
Distribution $P_{s}(s)$ of the infinite-time spread complexity rescaled by its mean, $s\equiv S_{K,\infty}/\braket{\ski}$, over disorder realisations in both the ergodic [panels (a), (b)] and MBL [panels (c), (d)] regimes, for different $L$. In the ergodic phase the distribution clearly approaches a Gaussian, 
 whereas in the MBL \new{regime} it is an exponential distribution, as indicated by the blue dashed lines denoting $e^{-s}$.
}
\label{fig:Sk-dist}
\end{figure}

We start with the distribution of the infinite-time spread complexity over disorder realisations; in particular the distribution $P_{s}(s)$ of the late-time complexity rescaled by its mean, $s = \ski/\braket{\ski}$,
numerical results for which are  shown in Fig.~\ref{fig:Sk-dist} for 
representative $W$'s in each of the ergodic and MBL \new{regimes}.

\begin{figure}
\includegraphics[width=\linewidth]{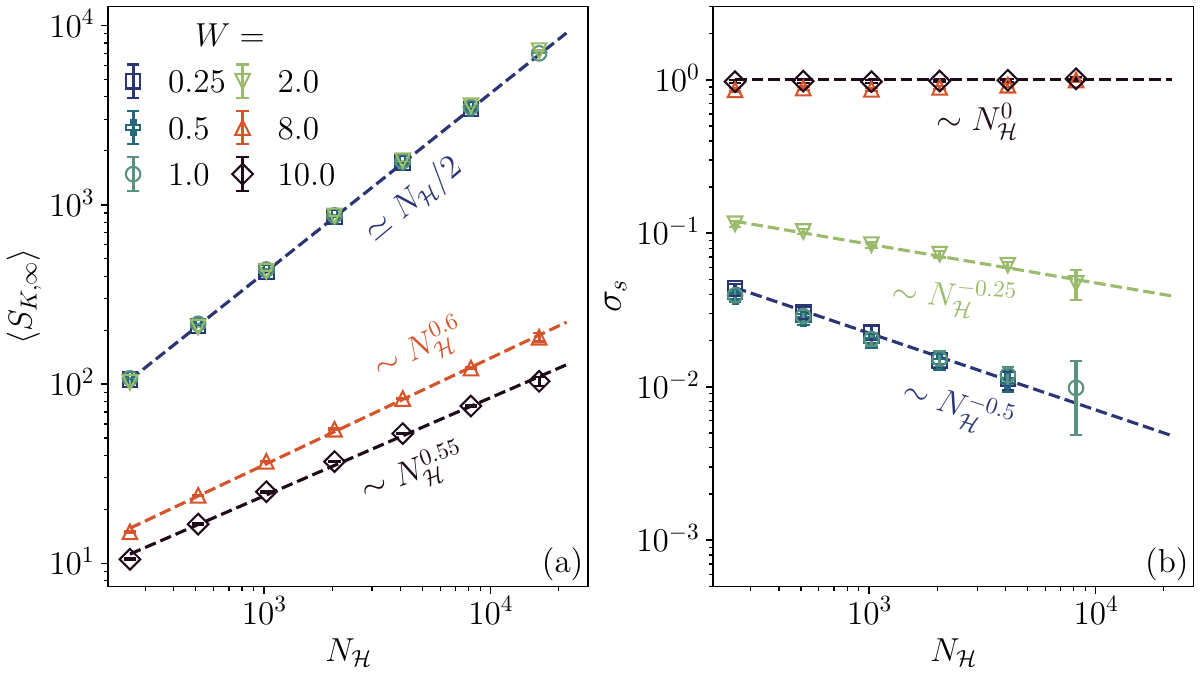}
\caption{
(a) Disorder averaged $\braket{S_{K,\infty}}$ as a function of $\nh$ on logarithmic scales, for different disorder strengths $W$. The straight line fits show that $\braket{S_{K,\infty}}\sim\nh^\alpha$ with $\alpha = 1$ in the ergodic \new{regime} and $\alpha<1$ in the MBL \new{regime}, as indicated explicitly next to fits.
(b) Scaling with $\nh$ of the standard deviation of $\ski$ relative to its mean, $\sigma_s \equiv {\rm std}[\ski]/\braket{\ski}$. The scaling exponents $\varsigma$, defined via $\sigma_s\sim\nh^{-\varsigma}$, are mentioned explicitly next to the fits. In the deep ergodic regime $\varsigma=1/2$, whereas in the MBL regime 
$\varsigma\simeq 0$.
}
\label{fig:SK-mean-std}
\end{figure}

In the ergodic regime the distributions are well described by a Gaussian with mean 1
\eq{
P_s^{\pd}(s) = \frac{1}{\sqrt{2\pi \sigma_s^2}}\exp\left[-\frac{(s-1)^2}{2\sigma_s^2}\right]\,,
\label{eq:P-ski-erg}
}
and clearly narrow rapidly with increasing system size. Results for the mean $\braket{\ski}$ are given in Fig.~\ref{fig:SK-mean-std}(a), from which it is evident that in the ergodic regime,
\eq{
\braket{\ski}\simeq \nh/2\,.
\label{eq:ski-mean-erg}
}
 Corresponding results for the standard deviation $\sigma_{s}$, given in
Fig.~\ref{fig:SK-mean-std}(b), show that it decays exponentially with system size as
\eq{
\sigma_s = \frac{{\rm std}[\ski]}{\braket{\ski}}\sim \nh^{-\varsigma}\,,
}
where $0<\varsigma<1$ decreases with increasing $W$. In particular, note that deep in the ergodic phase $\varsigma=1/2$. Since $\sigma_{s}$ is exponentially small in the system size $L$, $\lim_{L\to\infty}P_s(s)\to \delta(s-1)$ is thus
$\delta$-distributed in the thermodynamic limit. Recalling that $\ski$ is a measure of the size of the late-time wave function on the Krylov chain, the 
above results imply that in the ergodic regime the wave function is maximally spread out over the entire Krylov chain, with negligible fluctuations in the spread.

The observations above can be understood by considering the eigenvectors deep in the ergodic phase to be well represented by Gaussian random vectors. As such one 
considers the $\braket{k_n|E}\sim {\cal N}(0,\nh^{-1/2})$ to be uncorrelated Gaussian random numbers with zero mean and standard deviation $\nh^{-1/2}$; which
implies that $\Lambda_{n,\ket{E}} =|\langle k_{0}|E\rangle|^{2}|\langle k_{n}|E\rangle|^{2}$ (Eq.~\ref{eq:Lambda-n-E}) is a random number with mean $\nh^{-2}$ and standard deviation $2\sqrt{2}\nh^{-2}$ (for any $n>0$). Using this, it is easily shown that  $\Lambda_n =\sum_{n}\Lambda_{n,\ket{E}}$ is a random number with mean $\nh^{-1}$ and standard deviation $2\sqrt{2}\nh^{-3/2}$.
So from Eq.~\ref{eq:SK-inf}, 
\eq{
\ski = \sum_{n=0}^{\nh-1}Z_n^{\pd}\,
}
where $Z_{n}$ ($=n\Lambda_{n}$) is a random number with mean $n\nh^{-1}$ and standard deviation $2\sqrt{2}n\nh^{-3/2}$. The $Z_n$'s are not however independent,
as is explicit from the fact that $\braket{Z_nZ_m}\neq\braket{Z_n}\braket{Z_m}$ for $n\neq m$. More specifically, under the assumption above that $\braket{k_n|E}\sim {\cal N}(0,\nh^{-1/2})$, it can be shown that for $n\neq m$ 
\eq{
\braket{Z_n^{\pd} Z_m^{\pd}} - \braket{Z_n^{\pd}}\braket{Z_m^{\pd}} 
= 2nm/\nh^3\,.
}
The covariance matrix for the random variables $\{Z_n\}$ therefore has the form
\eq{
\left[C_Z\right]_{nm}^{\pd}=\delta_{nm}^{\pd}\frac{8n^2}{\nh^3} + (1-\delta_{nm}^{\pd})\frac{2 nm}{\nh^3}\,.
}
Given that $\ski$ is a sum of $\nh$ correlated random numbers, it follows that for large $\nh$ it is Gaussian distributed  (invoking a multivariate Lyapunov version of the central limit theorem); as indeed is corroborated by the numerical
results in Fig.~\ref{fig:Sk-dist}[(a) and (b)]. The mean of this distribution is given by
\eq{
\braket{\ski} = \sum_{n=0}^{\nh-1}\braket{Z_n^{\pd}} = \frac{\nh-1}{2}\overset{\nh\gg 1}{\sim}\frac{\nh}{2}\,,
\label{eq:mean}
}
as also found numerically in Fig.~\ref{fig:SK-mean-std}(a).
The corresponding variance is given by
\eq{
{\rm var}[\ski] = \sum_{n,m}\left[C_Z\right]_{nm} &= \frac{(\nh^2+3\nh-2)(\nh-1)}{2\nh^2}\nonumber\\
\Rightarrow {\rm var}[\ski]&\overset{\nh\gg 1}{\sim}\frac{\nh}{2}\,.
}
This implies that in the ergodic regime,  $\sigma_s\sim\nh^{-1/2}$ which is indeed 
exactly the result found numerically in Fig~\ref{fig:SK-mean-std}(b) for sufficiently weak disorder. It is worth reiterating the conceptual point that, even though the eigenvector amplitudes on the Krylov chain in the weak-disorder ergodic regime can be considered to be uncorrelated random variables, the $\Lambda_n$'s (or equivalently the $Z_n = n\Lambda_n$) are in fact correlated for different $n$. This correlation is rooted in the presence of the common factors of $|\braket{E|k_0}|^{2}$in both $\Lambda_n$ and $\Lambda_m$ (see Eq.~\ref{eq:Lambda-n-E}) even for $n\neq m$.

The physical content of Eq.~\ref{eq:mean} is also quite transparent.
Recall that $\braket{\ski}$ gives the mean position on the Krylov chain,
as $t\to \infty$, following initiation of the system at the end of the chain 
(in $\ket{k_{0}}$). In the ergodic phase, with eigenstates delocalised over the chain, the long-time mean position should thus be the midpoint of the chain
(the length of which is $\nh$); i.e.\
$\braket{\ski}=\nh/2$, as in Eq.~\ref{eq:mean}.\\

The situation in the MBL regime is completely different. The distribution of 
$\ski$, or equivalently $P_s$, is qualitatively different from that in the ergodic regime. Results for the mean $\braket{\ski}$ in Fig.~\ref{fig:SK-mean-std}(a) show that 
\eq{
\braket{\ski}\sim \nh^{\alpha}\,
~~~~:~ \alpha<1\,
\label{eq:ski-mean-mbl}
}
with an exponent $\alpha <1$, \new{which is in fact significantly less that $1$ for the $W$'s considered and which decreases with increasing $W$.
Note that the data in Fig.~\ref{fig:SK-mean-std}(a) fall on almost perfect straight lines on the logarithmic axes, indicating the robustness of the $\alpha$ exponent.
To further confirm this, we also checked the local slopes with varying $L$ in the data, and found that there is little variation in them and that they converge well to the values reported.}
The ratio $\braket{\ski}/\nh \propto \nh^{\alpha -1}$ gives the fraction of the Krylov-chain length which is accessed in the long-time limit
following initiation of the system in $\ket{k_{0}}$. Since $\alpha <1$, it thus vanishes exponentially in system size $L$, which is of course symptomatic of the many-body localised character of the eigenstates.

In marked contrast to its  Gaussian counterpart in the ergodic phase, the
distribution  of $s=\ski/\braket{\ski}$ is very well described by an exponential form
\eq{
P_s^{\pd}(s) = e^{-s}\,,
\label{eq:p-ski-mbl}
}
as seen in Fig.~\ref{fig:Sk-dist}[(c), (d)]. This is significant, as it implies that the standard deviation of $\ski$ scales in the same way with $\nh$ as the mean,
\eq{
{\rm std}[\ski]\sim \nh^{\alpha}\,.
}

The net physical import of the above results is the following. In the ergodic regime $\ski$ is sharply distributed  (becoming a $\delta$-function as $L\to\infty$) around its mean, which scales $\sim \nh$, reflecting the ergodic and extended nature of the eigenstates on the Krylov chain. By contrast, in the MBL regime $\ski\sim\nh^{\alpha}$ with $\alpha<1$, reflecting the non-ergodic nature of the eigenstates. In addition, the distribution of $\ski$ is rather broad, as the mean and standard deviation scale in the same way with $\nh$; as we will see in 
Sec.~\ref{sec:stats-spread}, this is related to a large-deviation  character of the contributions to $\ski$ from different eigenstates.

\subsection{Profile of $\Lambda_n$ on Krylov chain \label{sec:lam-n}}

The quantity $\Lambda_{n}$ (Eq.~\ref{eq:Lambda-n-E}) gives the long-time probability $|c_{n}(t$$\to$$\infty)|^{2}$ (Eqs.~\ref{eq:cn(t)},\ref{eq:SK(t)}) that the system will be found on orbital $n$ of the Krylov chain, given its initiation in the $n=0$ orbital. While $\ski =\sum_{n}n\Lambda_{n}$ is itself the 
first moment of this distribution,  the profile of the wave function, encoded in $\Lambda_n$, clearly contains finer information about the anatomy of the state on the Krylov chain. It is to this that we now turn.

\begin{figure}
\includegraphics[width=\linewidth]{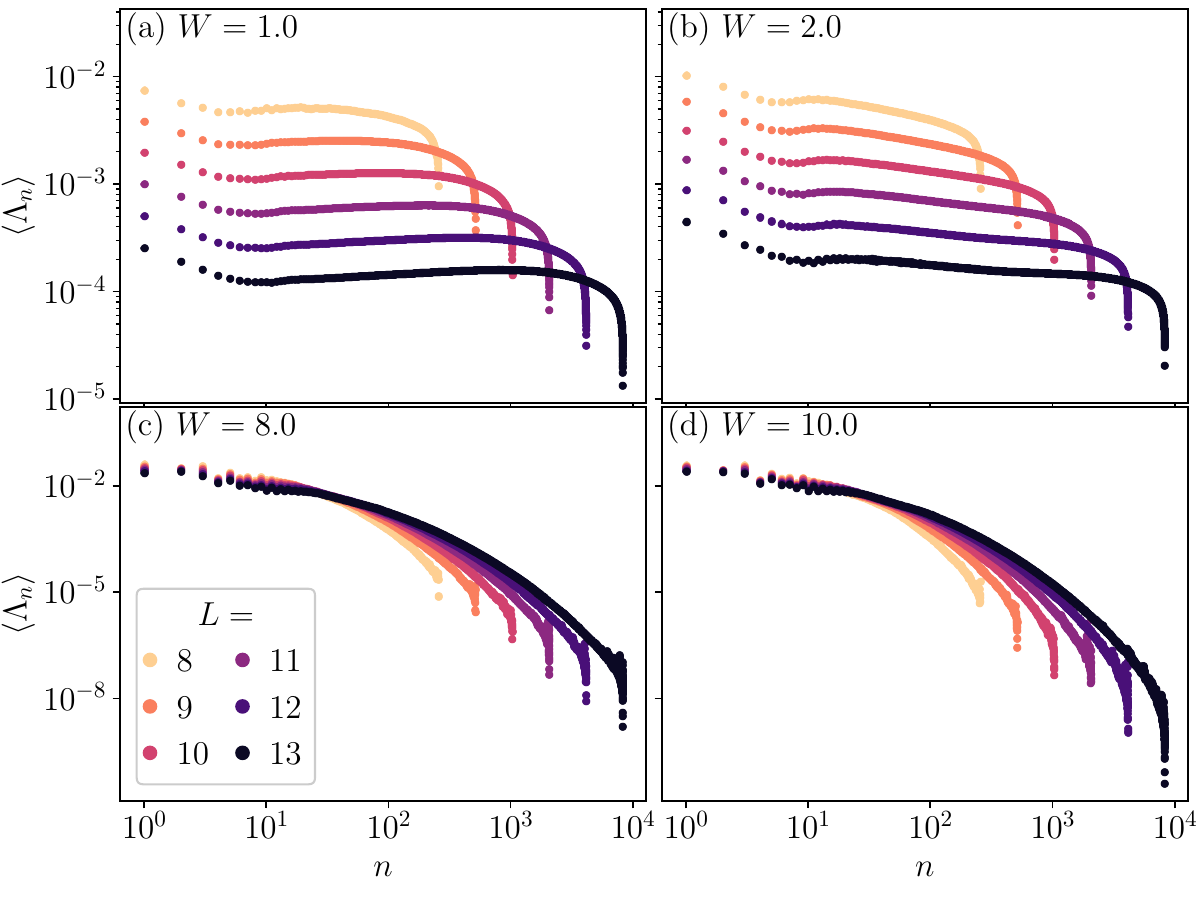}
\caption{Profile of the disorder-averaged $\braket{\Lambda_n}$ (defined in Eq.~\ref{eq:Lambda-n-E}) on the Krylov chain, for different $L$. Panels (a) and (b) correspond to the ergodic \new{regime} for $W=1,2$,  while panels (c) and (d) correspond to the MBL 
\new{regime} for $W=8, 10$.}
\label{fig:Lambdan-profile}
\end{figure}

\begin{figure}
\includegraphics[width=\linewidth]{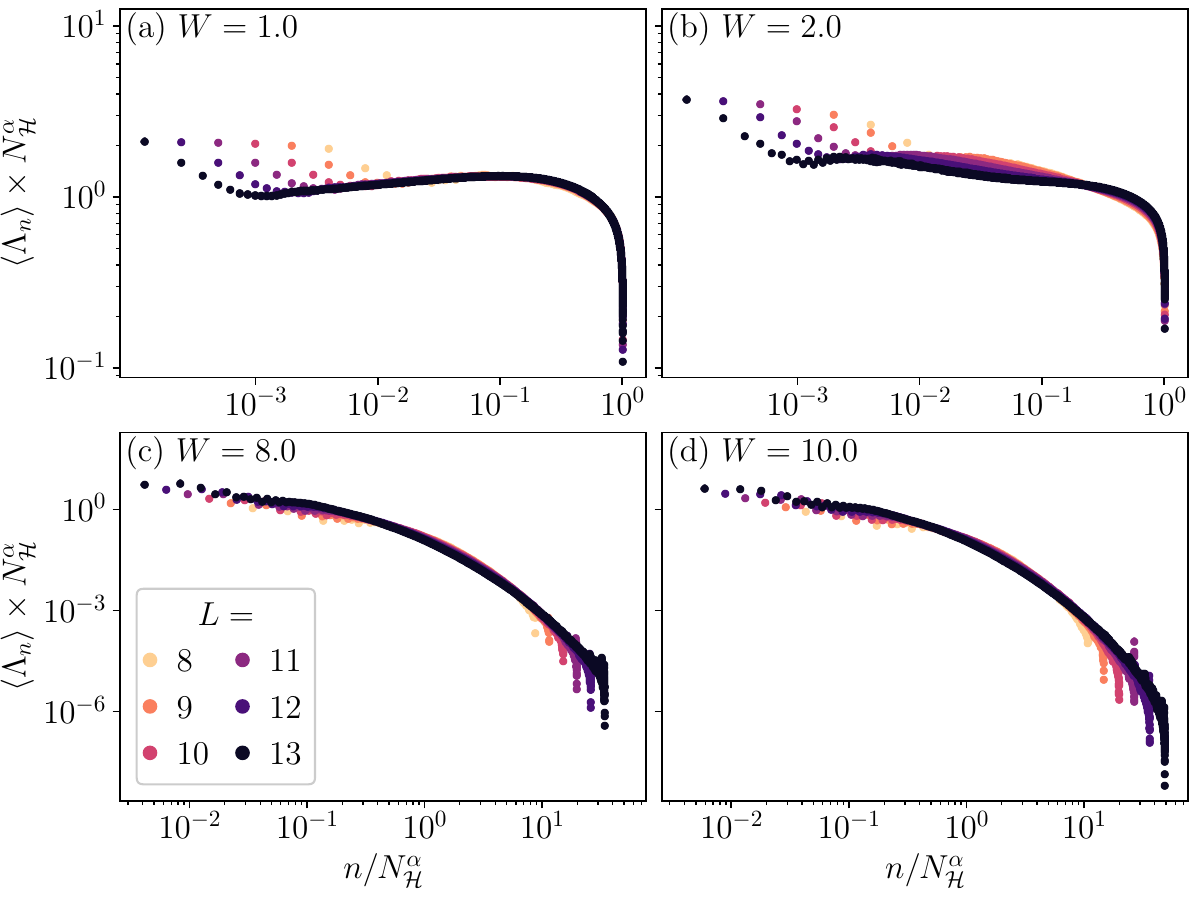}
\caption{Scaling of the $\braket{\Lambda_n}$ profile. $\braket{\Lambda_{n}}$$\times$$\nh^{\alpha}$ is plotted {\it{vs}} $n/\nh^{\alpha}$ $(\equiv x)$, with the exponents $\alpha$ obtained from the scaling of $\braket{\ski}$ in 
Fig.~\ref{fig:SK-mean-std}. The data demonstrates the scaling form in Eq.~\ref{eq:Lambdan-scaling} (with $\beta =\alpha$). Panels (a) and (b) correspond to the ergodic 
\new{regime} for $W=1,2$, while panels (c) and (d) correspond to the MBL 
\new{regime} for $W=8,10$.}
\label{fig:Lambdan-profile-scaled}
\end{figure}

Fig.~\ref{fig:Lambdan-profile} shows results for the disorder-averaged
$\braket{\Lambda_n}$ {\it{vs}} $n$ ($\in [0, \nh-1]$), in both the ergodic regime 
(panels (a)-(b)) as well as the MBL regime (panels (c)-(d)). In the ergodic regime, following a `transient' behaviour for $n\leq L$, the profile is approximately a flat plateau with $n$, indicating the ergodicity of all eigenstates. Of course, the normalisation $\sum_{n=0}^{\nh-1}\Lambda_n=1$ implies that the height of this plateau decreases exponentially with $L$, since its support itself grows exponentially with $L$. In the MBL regime on the other hand, there is a systematic decay of $\braket{\Lambda_n}$ with $n$, but the decay is slower for larger $L$ which suggests that $\braket{\Lambda_n}$ decays with $n$ over scales which scale as $\nh^\beta$. This motivates the general scaling ansatz 
\eq{
\braket{\Lambda_n^{\pd}} = \nh^{-\beta}f\left(\frac{n}{\nh^\beta}\right)\,,
\label{eq:Lambdan-scaling}
}
with $\beta\leq 1$ and the equality expected to be satisfied in the ergodic phase.
For $\nh =2^{L} \gg 1$, this implies for $\braket{\ski}=\sum_{n}n\braket{\Lambda_{n}}$
\eq{
\braket{\ski} &= \nh^{-\beta}\int_0^{\nh} dn~ n~f\left(\frac{n}{\nh^\beta}\right)\nonumber\\
&\simeq\nh^\beta \int_0^\infty dx~xf(x)\,.
}
However the results in Eq.~\ref{eq:ski-mean-erg} and Eq.~\ref{eq:ski-mean-mbl} for the scaling of $\braket{\ski}$ with $\nh$ imply that $\beta=\alpha$.
Indeed, numerical results presented in Fig.~\ref{fig:Lambdan-profile-scaled} show that the data for $\braket{\Lambda_n}$ is in excellent agreement with the scaling ansatz in Eq.~\ref{eq:Lambdan-scaling} with $\beta =\alpha$.
This shows how the profile of the decay of $\braket{\Lambda_n}$ with $n$, and its dependence on $\nh$, is intimately connected to the scaling of $\braket{\ski}$ with 
$\nh$.

\begin{figure}
\includegraphics[width=\linewidth]{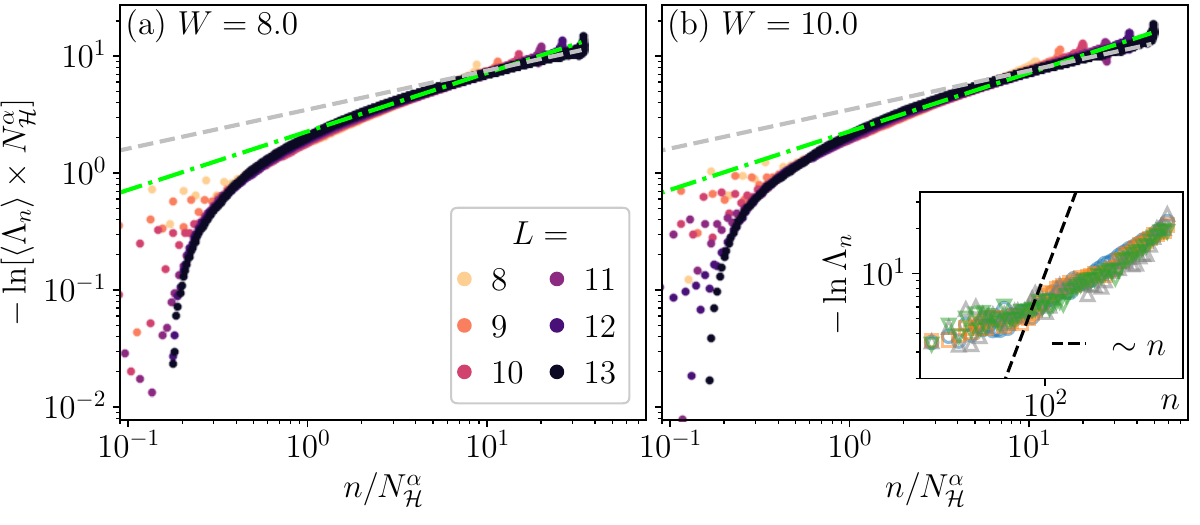}
\caption{Stretched-exponential behaviour of the scaling function for $\braket{\Lambda_n}$ (see Eq.~\ref{eq:Lambdan-scaling}) with $\beta=\alpha$ in the MBL \new{regime}
(and the $\alpha$ values those obtained in Fig.~\ref{fig:SK-mean-std}). Plots show the negative of the logarithm of the scaling function on logarithmic axes, where a straight line with slope $<1$ denotes a stretched exponential. The green dashed-dotted and the grey dashed lines correspond respectively to a stretch exponent of $\gamma=1/2$ and $1/3$ (see Eq.~\ref{fig:Lamban-stretch}).
Inset in (b) shows data for a few individual disorder realisations (different colours), each of which are also unambiguously stretched exponentials as the slope is clearly $<1$; the black dashed line denotes a slope equal to 1.}
\label{fig:Lambdan-profile-str-exp}
\end{figure}

We now consider the functional form of the scaling function, $f(x)$, in Eq.~\ref{eq:Lambdan-scaling}. In the ergodic phase, as discussed above, $f(x)\approx 1$ is an approximately flat function. On the other hand, as shown in panels (c) and (d) of Fig.~\ref{fig:Lambdan-profile-scaled}, in the MBL \new{case} $f(x)$ decays systematically with $n$. Note that on logarithmic axes, the latter data curves downwards, indicating a decay faster than any power law. This leads us to posit that the decay is exponential, possibly stretched. To confirm this, in Fig.~\ref{fig:Lambdan-profile-str-exp} we plot $-\ln[f(x)]$ as a function of $x$, on logarithmic axes. The results are consistent with $-\ln[f(x)] \sim x^{1/2}$, which in turn implies a stretched exponential decay of $\Lambda_n$, 
\eq{
\braket{\Lambda_n^{\pd}} \sim \nh^{-\alpha} \exp\left[-c\left(\frac{n}{\nh^\alpha}\right)^\gamma\right]\,,
\label{fig:Lamban-stretch}
}
where $\gamma\simeq 1/2$ (and $c>0$). 
While $\gamma\simeq 1/2$ is a good description of the stretched exponential for the bulk of the data, we note that for the larger values of $n/\nh^{\alpha}$, the data is better described by $\gamma=1/3$ (grey dashed lines in Fig.~\ref{fig:Lambdan-profile-str-exp}); we will return to this possibility shortly. While the precise value of the stretch exponent is slightly ambiguous within our numerical results, what stands robustly is the qualitative statement that $\braket{\Lambda_n}$ in the MBL regime has a stretched-exponential profile.

\subsection{Phenomenological theory}

In this section, we provide a simple phenomenological
theory for the numerical results presented above,
% when averaged over disorder realisations, 
specifically in the MBL regime. In particular, this provides a plausible rationalisation of the stretched exponential decay of $\braket{\Lambda_n}$ with $n$ as well as the exponential distribution of $S_{K,\infty}$ over disorder realisations. The two key ingredients that enter the phenomenological picture are: 
\begin{enumerate}
\item[] (i) For a given disorder realisation, there is a normalised distribution
over eigenstates, $P_{\xi_{|E\rangle}}(\xi)$, of lengthscales $\xi_{|E\rangle}$ 
on the Krylov chain, of form
\begin{equation}
\label{eq:31}
P_{\xi_{|E\rangle}^{\pd}}(\xi) =\frac{1}{\xibd}\times p_{\xi_{|E\rangle}}^{\pd}\left(\xi/\xibd\right).
\end{equation}
The distribution has a characteristic lengthscale $\xibd$, which 
without loss of generality can be defined via its first moment,
$\xibd =\int d\xi ~\xi ~P_{\xi_{|E\rangle}}(\xi)$.
\item[] (ii) Over an ensemble of disorder realisations, 
 $\xibd$ itself has a distribution $P_{\xibd}(\xibd)$ of form 
\begin{equation}
\label{eq:32}
P_{\xibd}^{\pd}(\xibd)=\frac{1}{\zb}\times p_{\xibd}^{\pd}(\xibd/\zb),
\end{equation}
and is likewise characterised by a lengthscale defined by
$\zb = \int d\xibd~\xibd~P_{\xibd}(\xibd)$.
\end{enumerate}
The starting point of the phenomenological picture is the conjecture 
that, for a given disorder realisation,
\begin{equation}
\label{eq:33}
\begin{split}
\Lambda_{n}^{\pd} ~=&~\sum_{E}
|\langle k_{0}^{\pd}|E\rangle|^{2}|\langle k_{n}^{\pd}|E\rangle|^{2}
\\
~\simeq &~ 
\frac{1}{\mathcal{N}}\sum_{E} |\langle k_{0}^{\pd}|E\rangle|^{2} ~e^{-n/\xi_{|E\rangle}^{\pd}}
\end{split}
\end{equation}
with $\mathcal{N}$ chosen to ensure the normalisation $\sum_{n}\Lambda_{n} =1$ (Eq.~\ref{eq:Lambda-n-sum-rule}). The reason we posit a distribution of lengthscales 
$\xi_{|E\rangle}$ over eigenstates at the level of each realisation
is because, as demonstrated in the preceding section, exact diagonalisation
(ED) results (Fig.~\ref{fig:Lambdan-profile-str-exp}) show that $\Lambda_{n}$ exhibits stretched exponential behaviour for typical disorder realisations as well as for the disorder-averaged $\langle \Lambda_{n}\rangle$; so stretched exponential behaviour is not solely due to averaging over disorder realisations. 
Eq.\ \ref{eq:33} can be written as
\begin{equation}
\label{eq:34}
\Lambda_{n}^{\pd} =\frac{1}{\mathcal{N}} \int_{0}^{\infty}d\xi~ e^{-n/\xi}
P_{\xi_{|E\rangle}}^{\pd}(\xi)
\end{equation}
with $P_{\xi_{|E\rangle}}(\xi)$ given by
\begin{equation}
\label{eq:35}
P_{\xi_{|E\rangle}}^{\pd}(\xi)~=~
\sum_{E} |\langle k_{0}^{\pd}|E\rangle|^{2}~\delta\big(\xi-\xi_{|E\rangle}^{\pd}
\big).
\end{equation}
The form of the lower right hand  side of Eq.\ \ref{eq:33} for $\Lambda_{n}$ is
conceptually apt, as reflected in $|\langle k_{0}|E\rangle|^{2}$
appearing explicitly in the summand. So if $|\langle k_{0}|E\rangle|^{2}$ for some given eigenstate $|E\rangle$ is `negligibly small', then that term
in the summand makes a negligible contribution to $\Lambda_{n}$. That
is physically sensible, since only a vanishing fraction of eigenstates make a significant contribution to the eigenstate sum in the first line of Eq.\ \ref{eq:33}.\footnote{For example, a rough caricture of MBL eigenstates takes the $|\langle k_{0}|E\rangle|^{2}$ to be non-zero and of order $\sim \nh^{-D}$ ($D<1$) only for $\sim \nh^{D}$ eigenstates (such that $\sum_{E}|\langle k_{0}|E\rangle|^{2}=1$); so that only a vanishing fraction $\sim \nh^{D-1}$ of eigenstates contribute to $\Lambda_{n}$.}
Relatedly, that $|\langle k_{0}|E\rangle|^{2}$ appears in Eq.\ \ref{eq:35} 
 reflects that  $P_{\xi_{|E\rangle}}(\xi)$ is the distribution of $\xi$ for those
eigenstates which make a non-negligible contribution to $\Lambda_{n}$: the
$|\langle k_{0}|E\rangle|^{2}$s in effect control the fraction of eigenstates which can make a non-negligible contribution to the eigenstate sum in Eq.\ \ref{eq:35}.

From Eq.\ \ref{eq:34}, the normalisation $\sum_{n}\Lambda_{n} =1$ gives (for 
$\nh =2^{L} \gg 1$) that $\mathcal{N}\equiv\int_{0}^{\infty}d\xi ~P_{\xi_{|E\rangle}}(\xi)\int_{0}^{\infty}dn~ e^{-n/\xi}$ $=\int_{0}^{\infty}d\xi ~\xi~P_{\xi_{|E\rangle}}(\xi) =\xibd$. The profile of $\Lambda_{n}$ for a given disorder realisation can thus be expressed as 
\begin{equation}
\label{eq:36}
\Lambda_{n}^{\pd} =\frac{1}{\xibd} \int_{0}^{\infty}d\xi~ e^{-n/\xi}
P_{\xi_{|E\rangle}}^{\pd}(\xi)
\end{equation}
which, using the form  Eq.\ \ref{eq:31} for $P_{\xi_{|E\rangle}}(\xi)$, 
yields the scaling form
\begin{equation}
\label{eq:37}
\Lambda_{n}^{\pd} = \frac{1}{\xibd} \times g\left(n/\xibd\right)
\end{equation}
with $g$ a function solely of $n/\xibd$. The disorder-averaged 
$\langle \Lambda_{n}\rangle$ is correspondingly given by
\begin{equation}
\label{eq:38}
\langle \Lambda_{n}^{\pd}\rangle = 
\int_{0}^{\infty}d\xibd ~P_{\xibd}^{\pd}(\xibd)~\Lambda_{n},
\end{equation}
and hence via Eqs.\ \ref{eq:37},\ref{eq:32} is of form
\begin{equation}
\label{eq:39}
\langle \Lambda_{n}^{\pd}\rangle = \frac{1}{\zb} \times
f\left(\frac{n}{\zb}\right)
\end{equation}
where $f$ is a function solely of $n/\zb$. This is precisely the scaling 
behaviour found numerically (Eq.~\ref{eq:Lambdan-scaling} with $\beta =\alpha$),
 viz.\ $\langle \Lambda_{n}\rangle= \nh^{-\alpha}f(n/\nh^{\alpha})$.
It also identifies the Krylov chain localisation length 
$\zb \propto \nh^{\alpha}$; which, while  exponentially large in the system size 
$L$, is a vanishing fraction of the chain length $\nh$ (as $\alpha <1$), again 
symptomatic of the many-body localised nature of the eigenstates.

Although the discussion above is independent of the precise forms of the distributions in Eqs.\ \ref{eq:31},\ref{eq:32}, we can infer a specific form for 
the distribution $P_{\xibd}(\xibd)$ of localisation lengths $\xibd$ over disorder realisations. From 
$S_{K,\infty}=\sum_{n} n\Lambda_{n} \equiv \int_{0}^{\infty}dn~n\Lambda_{n}$,
together with Eqs.\ \ref{eq:36},\ref{eq:31}, it is easily seen that
$S_{K,\infty}=u \xibd$ (with $u$ a constant), and hence the disorder-averaged 
$\langle S_{K,\infty}\rangle = u \langle \xibd\rangle =u \zb$;
such that 
\begin{equation}
s=\frac{\ski}{\braket{\ski}} = \frac{\xibd}{\zb}
\end{equation}
is equivalently just the ratio of $\xibd$ to its mean value $\zb$.
The distributions $P_{s}(s)$ and $P_{\xibd}(\xibd)$ are thus
related by $P_{\xibd}(\xibd)= (1/\zb)P_{s}\big(s=\xibd/\zb\big)$. 
The fact that $P_{s}(s) =e^{-s}$ is found from numerics to be an exponential distribution (see Fig.~\ref{fig:Sk-dist} and Eq.~\ref{eq:p-ski-mbl}) in turn implies that $P_{\xibd}(\xibd)$ is also an exponential distribution, specifically
\begin{equation}
\label{eq:40}
P_{\xibd}^{\pd}(\xibd)= \frac{1}{\zb}~\exp\left[-\xibd/\zb
\right].
\end{equation}
Motivated by this, we assume also the simplest such exponential
distribution for $P_{\xi_{|E\rangle}}(\xi)$, 
\begin{equation}
\label{eq:41}
P_{\xi_{|E\rangle}}^{\pd}(\xi)=
\frac{1}{\xibd}~\exp\left[-\xi/\xibd
\right].
\end{equation}
With this, Eq.\ \ref{eq:36} gives explicitly 
\begin{equation}
\label{eq:42}
\Lambda_{n}^{\pd} =
\frac{2}{\xibd} \sqrt{\frac{n}{\xibd}}~
K_{1}^{\pd}\left(2 \sqrt{\frac{n}{\xibd}}
\right),
\end{equation}
with $K_{1}$ a modified Bessel function of the second kind.
From the large-$x$ asymptotics $K_{1}(x) \propto e^{-x}$, 
Eq.\ \ref{eq:42} shows that $\Lambda_{n}$ indeed has a stretched 
exponential decay, with a stretch exponent of $\gamma =1/2$.
However, we still need to average the profile over disorder realisations
as in Eq.\ \ref{eq:38}, and from Eqs.\ \ref{eq:40},\ref{eq:42} this 
yields
\begin{equation}
\label{eq:43}
\langle \Lambda_{n}^{\pd}\rangle ~=~
\frac{1}{\zb} y^{1/2}~\mathcal{I}(y) ~~~~~~:~y=4n/\zb
\end{equation}
where $\mathcal{I}(y)=2\int_{0}^{\infty}du~ e^{-1/u^{2}}K_{1}(\sqrt{y} u)$.
While the explicit form of this integral (a Meijer G-function) is not illuminating, its large-$y$ asymptotics are. In particular, it  can be shown that
\begin{equation}
\label{eq:44}
\frac{\ln\big[ -\ln \mathcal{I}(y)\big]}{\ln y}
\sim \frac{1}{3} +\frac{\ln\big[3/2^{2/3}\big]}{\ln y}
+\frac{2^{-1/3}}{3} y^{-1/3} +\cdots
\end{equation}
which in turn implies that the stretch exponent for the stretched exponential decay of $\langle \Lambda_{n}\rangle$ is in fact $\gamma =1/3$. As discussed in the preceding section (\ref{sec:lam-n}) with regard to 
Fig.~\ref{fig:Lambdan-profile-str-exp}, this is also consistent with
the numerical data, particularly at large values of 
$n/\zb \propto n/\nh^{\alpha}$. That being said, the convergence of
$\gamma$ to its asymptotic value of $1/3$ is extremely slow, as evident
from the subleading corrections in Eq.\ \ref{eq:44}. In fact, the full
$\ln[-\ln\mathcal{I}(y)]/\ln y$ throughout the regime $10 \lesssim y \lesssim 100$ -- as relevant to the numerical results shown in Fig.~\ref{fig:Lambdan-profile-str-exp} --lies within $\sim 0.05$ of $1/2$. This offers a possible rationale for the numerical observation of $\gamma \simeq 1/2$ over the range of practically accessible system sizes, but which ultimately tends asymptotically to $1/3$.

\section{Statistics of eigenstate contributions to spread complexity
\label{sec:stats-spread}}

The infinite-time spread complexity can be written as a sum of contributions from individual many-body eigenstates. Specifically, Eq.~\ref{eq:SK-inf} may be recast as
\eq{
\ski = \sum_E \ske \,,
}
where
\eq{
\ske = \sum_n n\, \Lambda_{n,\ket{E}}^{\pd}
=\sum_n n~|\langle k_{0}^{\pd}|E\rangle|^{2}|\langle k_{n}^{\pd}|E\rangle|^{2}
}
defines the Krylov complexity associated with the eigenstate $\ket{E}$. 
This decomposition naturally leads to the question of which eigenstates dominate the sum, and what statistical properties of $\ske$ control $\ski$.

One expects eigenstates with anomalously large $\ske$ to correspond to wavefunctions whose support on the Krylov chain extends to atypically high Krylov orbitals, signalling the presence of long-ranged resonances on the Krylov chain. 
The central question we address in this section is whether such atypical eigenstates exist and, if so, how their statistics imprint themselves on the infinite-time spread complexity $\ski$; more precisely, whether $\ski$ is built from a finite fraction of eigenstates, or is instead dominated by a vanishing fraction residing in the extreme tails of the distribution of $\ske$.

\begin{figure}
\includegraphics[width=\linewidth]{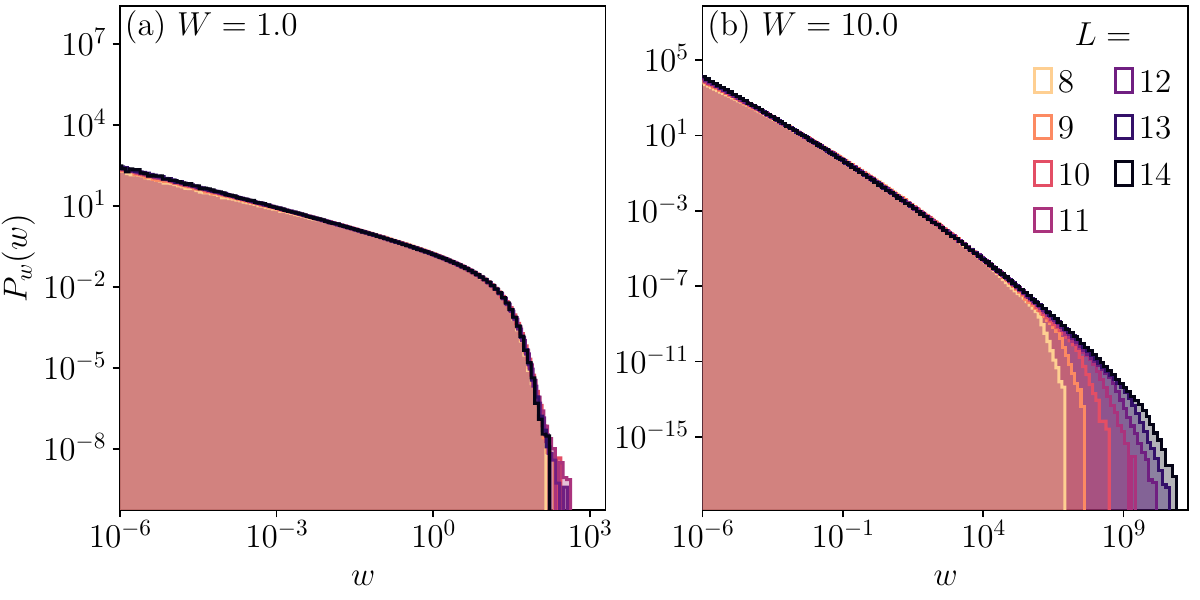}
\caption{Probability distribution $P_{w}(w)$ of $w_{\ket{E}}=S_{K,\ket{E}}/S_{K,{\rm typ}}$, over eigenstates $\ket{E}$ and disorder realisations. Panel (a) 
shows results for the ergodic phase at $W=1$, and panel (b) for the MBL 
\new{regime} at $W=10$. The data shows the development of increasingly longer tails in $P_w(w)$ with increasing $L$ in the MBL \new{regime}, but not in the ergodic \new{regime}. This is highlighted 
by the remarkably large difference in the scales of the support of $P_w(w)$.
}
\label{fig:Pw}
\end{figure}

To first obtain a broad brush view we define
\eq{
w_{\ket{E}}^{\pd} = \ske^{\pd}/S_{K,{\rm typ}}^{\pd}\,,
}
where $w_{\ket{E}}$ is the complexity of the eigenstate relative to the typical eigenstate complexity defined via 
$\ln S_{K,{\rm typ}} =\nh^{-1}\braket{\sum_{E}\ln\ske}$. 
Numerical results for the distribution of $w_{\ket{E}}$ over both eigenstates and disorder realisations,
$P_{w}(w)=\nh^{-1}\langle \sum_{E}\delta(w - w_{\ket{E}})\rangle$,
are shown in Fig.~\ref{fig:Pw}. From this it is evident that in the ergodic 
\new{regime}, the distribution is converged with $L$ and well behaved with finite support.
By contrast, in the MBL \new{regime} the distribution is significantly broader and develops longer tails with increasing system size.  This is the first indication that $\ski$ in the MBL \new{regime}  is indeed dominated by a few eigenstates whose $\ske$ lie in the tails of $P_w$.

To characterise this quantitatively, we introduce the generalised moments
\eq{
\skiq = \sum_E \ske^q \,,
\label{eq:Sqinf-sum}
}
which allow for a large-deviation analysis analogous to that on the Fock-space 
graph~\cite{biroli2024largedeviation,miranda2025large}. 
Defining $\ze = \ln \ske$, the disorder-averaged moments can be written as
\eq{
\braket{\skiq} = \int dz~P_z^{\pd}(z)\, e^{q z} \, e^{L \ln 2} \,,
\label{eq:Sqinf-integ}
}
where $P_z(z) = \nh^{-1}\braket{\sum_E\delta(z-z_{\ket{E}})}$  is the distribution of $z_{\ket{E}}$ over eigenstates and disorder realisations.

\begin{figure}
\includegraphics[width=\linewidth]{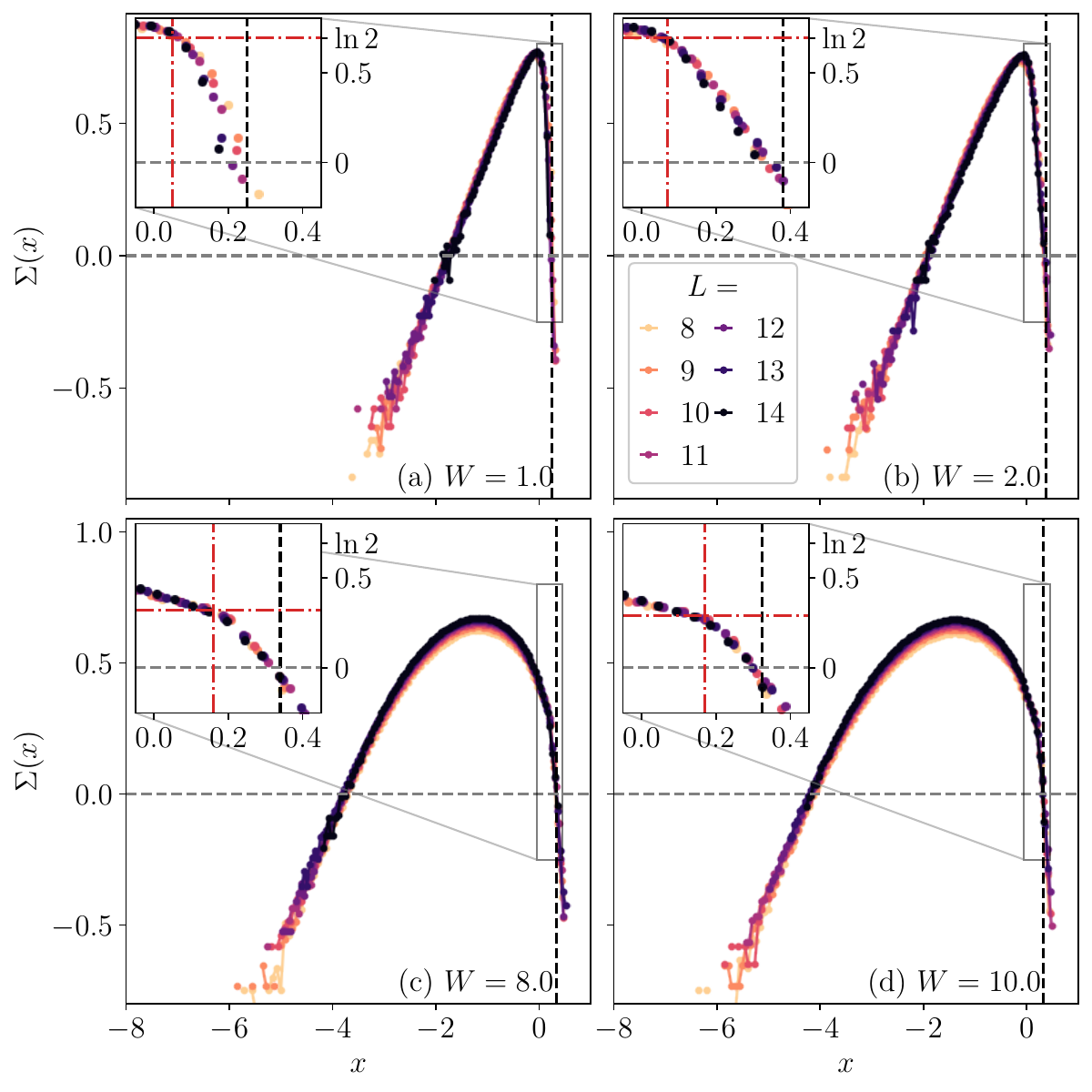}
\caption{$\Sigma(x)$ defined in Eq.~\ref{eq:Sigmax} for disorder strengths  in the ergodic \new{regime} [panels (a),(b)] and MBL \new{regime} 
[panels (c),(d)]. The $x$-domain where $\Sigma(x)>0$ (i.e.\ above the grey, horizontal dashed line) is the physically valid regime for the saddle point to lie. The black vertical dashed line denotes the value of $x$ at the upper extremity of this regime.
Insets show a zoom of the region near the saddle point, as indicated by the grey boxes. In each inset, the red vertical line denotes the location of the saddle point $x_\ast$ whereas the red horizontal line marks the value $\Sigma(x_\ast)$. In the ergodic phase [panels (a),(b)], $\Sigma(x_\ast)\simeq \ln 2$ whereas in the MBL regime [panels (c),(d)], $\Sigma(x_\ast)<\ln 2$.}
\label{fig:Sigmax}
\end{figure}

$\ske$ is typically expected to scale polynomially with $\nh$, 
and thus $\ze$ to scale extensively with $L$. It is therefore convenient to introduce the intensive variable $x = z/L$, in terms of which 
Eq.~\ref{eq:Sqinf-integ} becomes
\eq{
\braket{\skiq} = \int dx~P_x^{\pd}(x)\, e^{L \ln 2} \, e^{L q x} \,,
\label{eq:Sqinf-integx}
}
where $P_x(x) = L P_z(Lx)$. The quantity
\eq{
N_x^{\pd} = P_x^{\pd}(x) e^{L \ln 2}\,
}
represents the total number density of eigenstates contributing a weight $e^{L q x}$ to $\skiq$. This naturally defines a corresponding entropy density
\eq{
\Sigma(x) = \frac{1}{L} \ln N_x^{\pd} = \ln 2 + \frac{1}{L} \ln P_x^{\pd}(x)\,,
\label{eq:Sigmax}
}
such that 
\eq{
\braket{\skiq} = \int dx~\exp\big(L\left[\Sigma(x) +q x
\right]
\big) \,.
\label{eq:Sqinf-integxsig}
}
Numerical results for $\Sigma(x)$, for representative disorder strengths in both the ergodic and MBL regimes, are shown in Fig.~\ref{fig:Sigmax}.
The important point to note is that $\Sigma(x)$ is independent of $L$, so is indeed a bona fide, intensive entropy density. This naturally suggests that the integral in Eq.~\ref{eq:Sqinf-integxsig} will be governed by a saddle-point structure.

Since $\Sigma(x)$ corresponds to an effective entropy density, physical consistency requires $\Sigma(x) > 0$, i.e.\ for the saddle-point solution to be physical, it must satisfy $\Sigma(x_\ast)>0$.
The saddle point $x_{\ast} \equiv x_\ast(q)$ governing Eq.~\ref{eq:Sqinf-integxsig}
is determined by
\eq{
q + \partial_x \Sigma(x)\vert_{x=x_\ast(q)} = 0 \,.
\label{eq:saddle-point}
}
If the solution lies in a regime where $\Sigma(x_\ast) < 0$, the saddle-point approximation breaks down and the integral becomes dominated by the boundary value $x_0$ defined by $\Sigma(x_0)=0$. This scenario is mathematically analogous to the freezing transition in directed polymers in random 
media~\cite{derrida1988polymers}. When $\Sigma(x_\ast) > 0$, such that the saddle-point analysis is applicable, note that if $\Sigma(x_\ast) < \ln 2$ then the number of contributing eigenstates scales with $\nh$ as
\eq{
N_{x_\ast}^{\pd} \sim e^{L \Sigma(x_\ast)} \sim \nh^{c}, \qquad c<1
}
with a power $c<1$, implying that only a vanishing fraction of the spectrum contributes. This formalism shows how the behaviour of $\Sigma(x)$ and the saddle point $x_\ast$, underpins the answer to the question of whether or not $\ski$ is dominated by the $\ske$ of a few eigenstates. In the following, we perform this analysis numerically.

From the numerically obtained probability distribution of $\ske$, we construct the distribution of $z_E=\ln \ske$ and hence the intensive variable $x=z_E/L$, yielding the distribution $P_x(x)$. Using Eq.~\ref{eq:Sigmax}, the entropy density $\Sigma(x)$ is then computed. This is shown in Fig.~\ref{fig:Sigmax}, for both the ergodic and MBL regimes. The data exhibit good convergence with system size, justifying the saddle-point analysis of Eq.~\ref{eq:Sqinf-integxsig}.
The values of $x$ at which $\Sigma(x)$ crosses zero, delimit the range in which the saddle-point solution remains valid.

From the $\Sigma(x)$ so derived, we take a numerical derivative and extract the saddle point $x_\ast(q)$ using Eq.~\ref{eq:saddle-point}. The key point to note from Fig.~\ref{fig:Sigmax} is that the saddle point lies well within the region where $\Sigma(x)>0$, for all disorder strengths considered, implying that $\ski$ is built from an exponentially large number of eigenstates. However, the nature of this contribution differs sharply between phases. In the ergodic \new{regime} we find $\Sigma(x_\ast)\simeq \ln 2$, indicating that a finite fraction of the spectrum contributes; this is shown by the red dashed-dotted lines in the insets to panels (a) and (b) in Fig.~\ref{fig:Sigmax}. By contrast, in the MBL \new{regime}, the value of $\Sigma(x)$ at the saddle point satisfies  $\Sigma(x_\ast) < \ln 2$ (see insets to panels (c) and (d) in Fig.~\ref{fig:Sigmax}), such that only a vanishing fraction of eigenstates contribute, despite their number still increasing exponentially with $L$.

This distinction is quantified by the scaling
\eq{
\braket{\skiq} \sim e^{L [q x_\ast(q) + \Sigma(x_\ast(q))]} \equiv \nh^{\alpha_q}\,,
}
with
\eq{
\alpha_q^{\pd} = \frac{q x_\ast(q) + \Sigma(x_\ast(q))}{\ln 2}\,.
}
Representative numerical values for $q=1$ are listed in Table~\ref{eq:ld-table} and are consistent with the direct scaling analysis 
of $\braket{\ski}$ shown in Fig.~\ref{fig:SK-mean-std}.

\begin{table}
\begin{center}
\begin{tabular}{|c|c|c|c|c|}
\hline
$W$&$x_\ast(q=1)$ & $\Sigma(x_\ast(q=1))$ & Estimated $\alpha_{q=1}$&$\alpha$ from $\braket{\ski}$\\
\hline
1 & 0.05 & 0.72 & 1.11 & 1.0\\
2 & 0.07 & 0.69 & 1.09 & 1.0\\
8 & 0.18 & 0.32 & 0.72 & 0.61\\
10 & 0.17 & 0.29 & 0.66& 0.57\\
\hline
\end{tabular}
\end{center}
\caption{Tabulating the location of the saddle point $x_\ast(q=1)$ and the corresponding value of $\Sigma(x_\ast)$, for disorder strengths in both the ergodic ($W=1,2$) and MBL ($W=8,10$) regimes. The fourth column shows the estimated value of $\alpha_{q=1}$ from the saddle-point analysis, while the rightmost column shows the exponent obtained directly from the scaling of 
$\braket{\ski}$ with $\nh$ (as in Fig.~\ref{fig:SK-mean-std}).}
\label{eq:ld-table}
\end{table}

A notable feature of the saddle-point structure is that $x_\ast$ in the MBL \new{regime} lies well to the right of the mode of $P_{x}(x)$, implying that $\ski$ is controlled by eigenstates with anomalously large $\ske$ compared to the bulk of the spectrum, and lying in the tail of its distribution.

\begin{figure}
    \includegraphics[width=\linewidth]{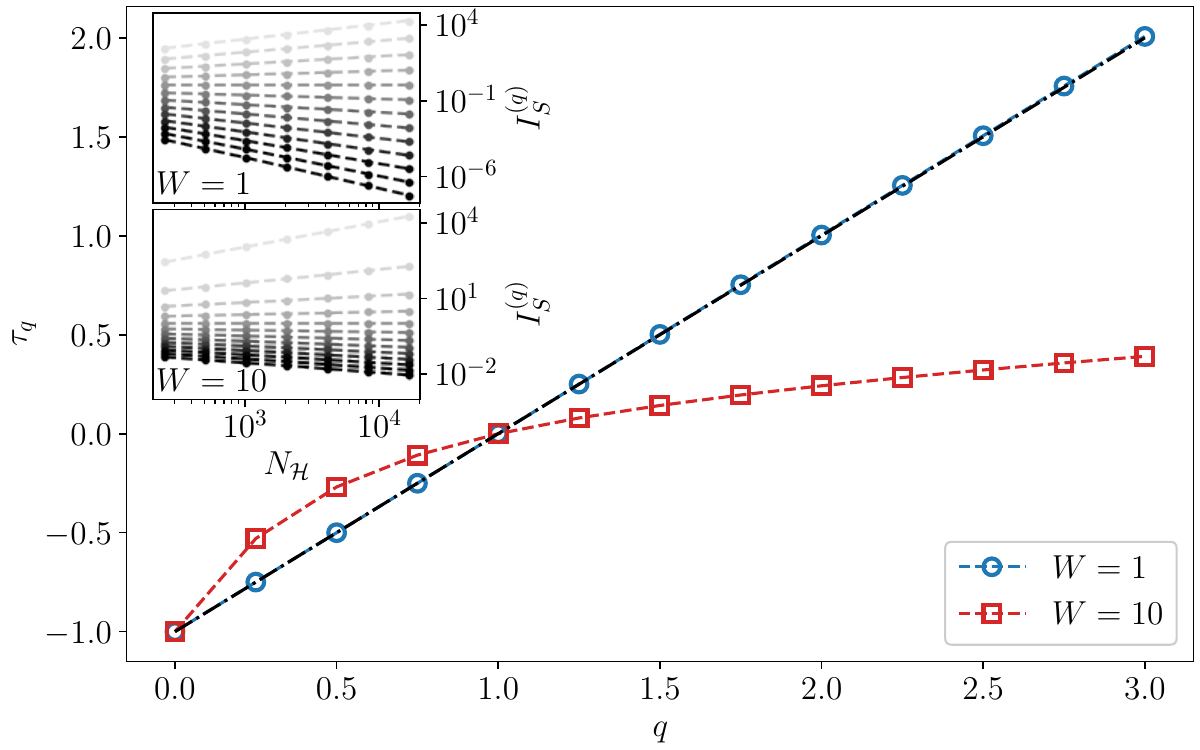}
    \caption{Scaling of the IPR of eigenstate complexities, defined in 
		Eq.~\ref{eq:SQIPR}. Main panel shows the scaling exponent $\tau_q$ 
	{\it{vs}} $q$ for $W=1$ and 10, representative of the ergodic and MBL regimes respectively. The black dashed-dotted line shows 		
		$\tau_q=q-1$, the expected result for full ergodicity.
		The non-linear $q$-dependence of $\tau_q$ for $W=10$ reflects 
multifractality in the MBL \new{regime}. Insets show the raw data for $I_S^{(q)}$ as a function of $\nh$, for $q=0.25,0.5,\cdots,3$ (darker colours  denote larger $q$) on logarithmic axes, straight-line fits to which were used to extract the $\tau_q$ values shown in the main panel.}
    \label{fig:Sq_IPR}
\end{figure}

Finally, the phenomenology described above is further corroborated by the inverse participation ratios (IPR) of the $\ske$ over all the energy eigenstates. Formally, the generalised $q^{\rm th}$-IPR is defined as 
\eq{
I_{S}^{(q)} = \left\langle\sum_E \left(\frac{\ske}{\sum_E\ske}\right)^q\right\rangle\,
\label{eq:SQIPR}
}
(with $\braket{\cdots}$ the disorder average), and is expected to scale with $\nh$ as 
\eq{
I_{S}^{(q)} \sim \nh^{\tau_{q}}~~
\mathrm{with}~
\tau_{q}^{\pd} = D_{q}^{\pd}(q-1)\,.
\label{eq:SQIPRTAU}
}
In the ergodic phase one expects $D_q=1$ for all $q$, while in the MBL phase
one expects  $D_q$ to be a non-trivial function of $q$ with $0<D_q<1$.
The numerical results shown in Fig.~\ref{fig:Sq_IPR} for $W=1$ and $10$
exemplify perfectly the above expectations.  The behaviour $\tau_{q}=(q-1)$ is clearly seen for the weak-disorder case  $W=1$, while the non-linear $q$-dependence of $\tau_{q}$ for $W=10$ indicates the multifractality characteristic of the MBL \new{regime}. This multifractal scaling of $I_{S}^{(q)}$ provides additional evidence for the fact that in the sum $\ski=\sum_E\ske$, only a vanishing fraction of the eigenstates -- but nevertheless an exponentially large (in $L$) number of them -- contribute.

\section{Concluding remarks \label{sec:conclusion}}
In this work we have analysed the anatomy and complexity of quantum states in Krylov space, in the ergodic and MBL \new{regimes} of a disordered, interacting spin chain. 
Using the Krylov basis generated by the Hamiltonian from an initial state, we characterised the infinite-time complexity of the state by its Krylov spread complexity, which provides a basis-optimised measure of complexity. This long-time spread complexity 
was shown to  distinguish sharply between the two \new{regimes}: in the ergodic \new{case}, it scales linearly with the Fock-space dimension $\nh$, reflecting spreading over a finite fraction of the Krylov chain;  whereas in the MBL \new{regime} it grows sublinearly, $\propto \nh^\alpha$ with $\alpha<1$, implying confinement to a vanishing fraction of the chain.
Beyond this scaling distinction, the Krylov-space anatomy of the long-time state 
also reveals a clear structural signature of the MBL \new{regime}. 
The infinite-time state develops a stretched-exponential profile along the Krylov chain, which can be understood as arising from a broad distribution of exponential decay lengthscales across eigenstates. Consistently, a large-deviation analysis shows that while the ergodic phase receives contributions from almost all eigenstates, the complexity in the MBL \new{regime} is dominated by rare resonant eigenstates lying in the tails of the distribution of the eigenstate Krylov spread complexity. 

These results highlight Krylov space as a particularly transparent framework for analysing the structure and spreading of many-body states, where the effective one-dimensional geometry of the Krylov chain provides a simple yet rich perspective on quantum complexity in disordered systems; and where in particular the length of the Krylov chain -- which is equal to the Fock-space dimension $\nh$ of the original Hamiltonian -- serves as the natural length scale in terms of which to  understand the scaling of physical properties with system size.

While the present work has centred on the infinite-time spread complexity, a natural immediate question is to understand the temporal dynamics of the Krylov spread complexity, in particular what further insight it gives into the nature of the MBL \new{regime}.
\new{Preliminary results (see Appendix~\ref{app:dynamics}) already suggest a marked difference in the dynamics in the two regimes, with the spread complexity growing anomalously slowly in the MBL regime.}
Looking further afield, one interesting direction would be to understand the connections between eigenstate Krylov spread complexities and other measures of their complexity, such as their entanglement structure. It  is known that entanglement properties are encoded in Fock-space correlations beyond multifractality or the lack thereof~\cite{detomasi2020multifractality,roy2022hilbert}. Asking analogous questions on the Krylov space would certainly be a fruitful endeavour.

\begin{acknowledgements}
DEL is grateful for the hospitality of the International Centre for Theoretical Sciences (ICTS-TIFR), Bengaluru, where part of this work was undertaken. BP and SR acknowledge support from the Department of Atomic Energy under Project Nos. RTI4019 and RTI4013. SR acknowledges support from SERB-DST, Government of India, under Grant No. SRG/2023/000858, and by a Max Planck Partner Group grant between ICTS-TIFR, Bengaluru and MPIPKS, Dresden.
\end{acknowledgements}

\begin{figure}
\includegraphics[width=\linewidth]{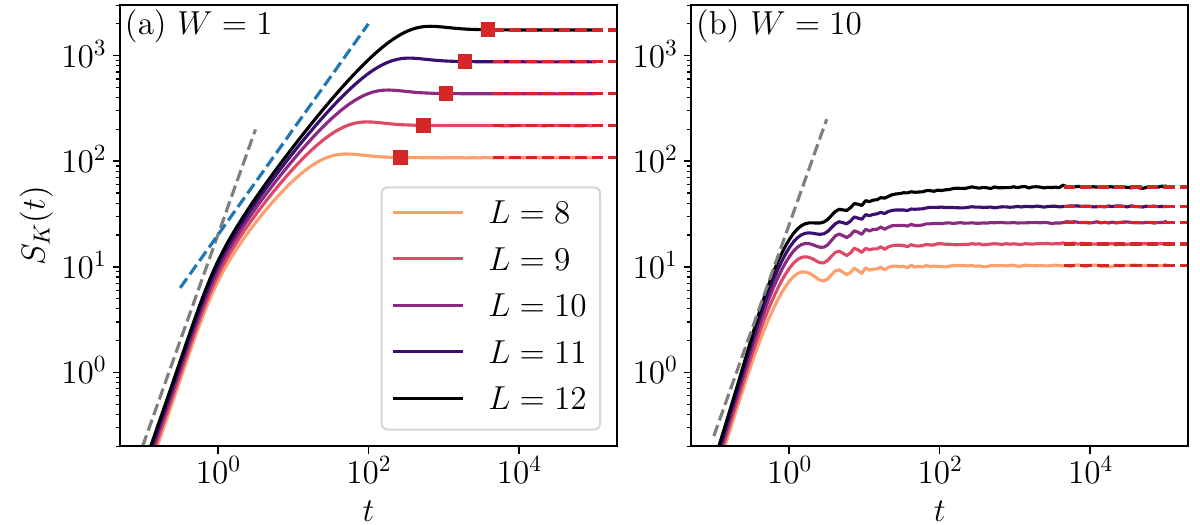}
\caption{\new{The temporal behaviour $S_K(t)$ {\it{vs}} $\Gamma t$ (with
$\Gamma \equiv 1$),
for (a) weak disorder ($W=1$)
and (b) strong disorder ($W=10$), for the TFI model Eq.~\ref{eq:ham-tfi} with system sizes $L$ as indicated;
shown as a log-log plot in the former case and a log-linear one for the
latter. In each case, the grey dashed line denotes the early time $\sim t^2$ behaviour. In the ergodic regime (panel a), the blue dashed line denotes the linear-in-time behaviour, and the red squares therein denote the Heisenberg time $t_H=\nh$. In both panels, the red dashed lines denote the $\ski$ values used in the scaling analysis of Fig.~\ref{fig:SK-mean-std} where the $t\to \infty$ limit was explicitly taken.}}
\label{fig:dyn-SK}
\end{figure}

\appendix
\new{
\section{Dynamics of the Krylov spread complexity \label{app:dynamics}}
In this appendix, we present briefly some basic numerical results for the temporal behaviour of $S_K(t)$, at both weak and strong disorder
(exemplified by $W=1$ and $10$ respectively).
These are shown in Fig.~\ref{fig:dyn-SK}. At the earliest times, there is a quadratic-in-time growth of the spread complexity, $S_K(t)\sim t^2$.
This follows quite trivially from perturbative arguments to leading order in powers of $\Gamma t$, which give 
\eq{
S_K(t)\overset{\Gamma t\ll 1}{\sim} t^2L\Gamma^2\,.
\label{eq:SK-smallt}
}
This leading low-$t$ behaviour is common to both disorder regimes, and extends
up to $\Gamma t \sim \mathcal{O}(1)$.
}

\new{In the weak-disorder, ergodic regime (Fig.~\ref{fig:dyn-SK}(a)), the early-time behaviour is followed by a universal linear-in-time growth of $S_K(t)$, indicative of ergodic transport; before crossing over to a late-time saturation value of $\sim\tfrac{1}{2}\nh$ at timescales on the order of the Heisenberg time $t_H\propto\nh$.
In the strong disorder, MBL regime by contrast (Fig.~\ref{fig:dyn-SK}(b)), the early-time behaviour is instead followed by an extremely slow growth of $S_K(t)$ towards its eventual saturation, which itself scales as $\nh^\alpha$ with  $\alpha<1$. 

Understanding in detail this rich dynamical behaviour remains an open question of obvious interest. In the present work, we have of course focussed on the long-time limit of the spread complexity, with $t\to \infty$ taken from the outset (as embodied in Eqs.~\ref{eq:SKinfdef}-\ref{eq:Lambda-n-E}) such that the resultant $\ski$ is determined by the 
character of the eigenstates of the Krylov Hamiltonian. We would however point out that this $\ski$ indeed emerges perfectly smoothly from the underlying dynamics: the red dashed lines in Fig.~\ref{fig:dyn-SK}, for both disorder strengths, show the $\ski$ values used in the scaling analysis of Fig.~\ref{fig:SK-mean-std},
which evidently agree with the long-time saturation values of $S_{K}(t)$.
}

%%%%%%%%%%%%%%%%%%%%%%%%%%%%%%%%%%%%%%%%%%%%%%%
\section{Some statistical properties of the Krylov Hamiltonian \label{app:anbn}}

\begin{figure*}
\includegraphics[width=\linewidth]{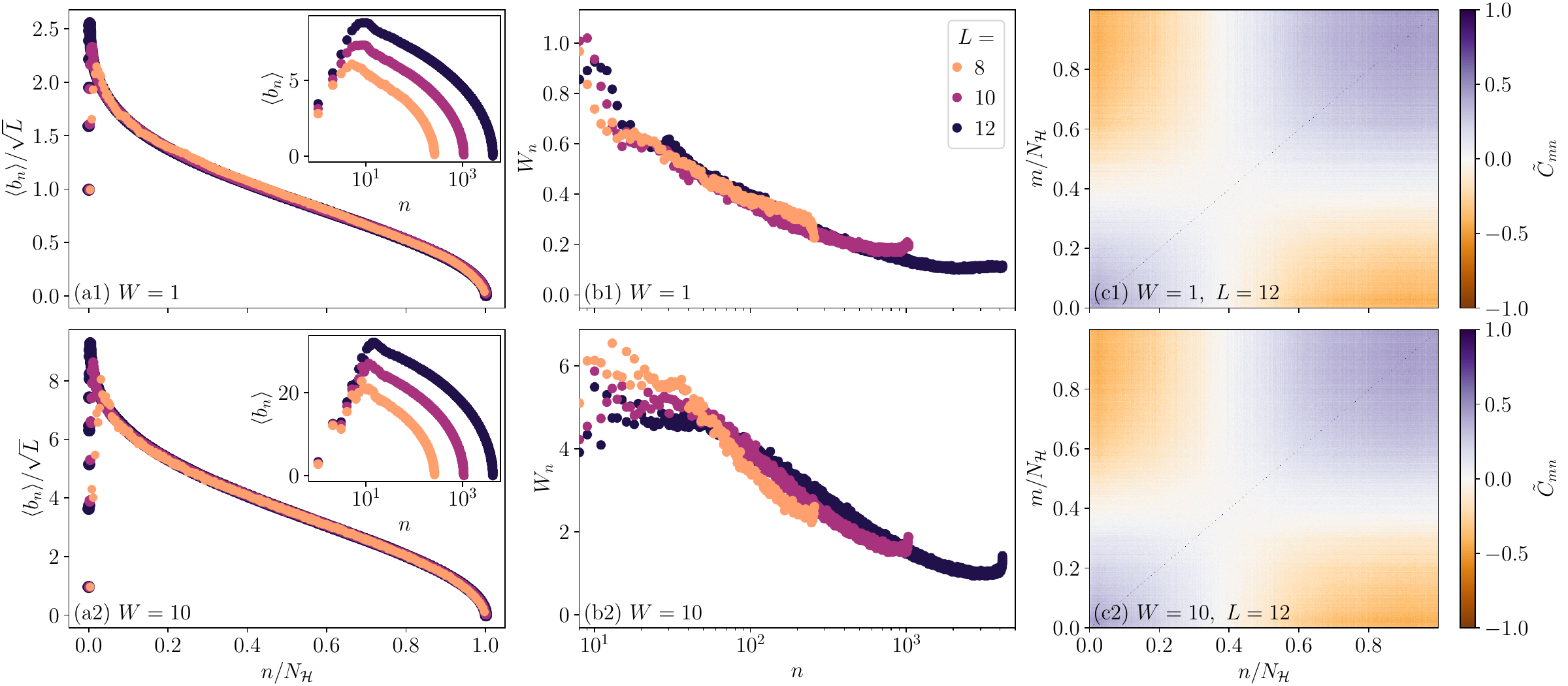}
\caption{(a1,a2) Scaling of the averaged hopping matrix element $\braket{b_n}$ in the Krylov Hamiltonian Eq.~\ref{eq:H-krylov-tridiag}, for $W=1$ (a1) and $W=10$ (a2), with system sizes $L$ as indicated. 
Insets show $\braket{b_{n}}$ {\it{vs}} $n$. Main panels show 
$\braket{b_{n}}/\sqrt{L}$ {\it{vs}} $n/\nh$, and demonstrate the scaling
form Eq.~\ref{eq:bn-scaling}, holding for both ergodic and MBL regimes.
(b1,b2) Profile of the effective disorder $W_n$ (defined in Eq.~\ref{eq:Wn-def})  
{\it{vs}} $n$, in the ergodic and MBL regimes.
(c1,c2) Correlation in the onsite disorder on the Krylov chain, defined in Eq.~\ref{eq:an-corr}, and shown as a heatmap. Data is for $L=12$.}
\label{fig:krylov-ham}
\end{figure*}

Here we discuss briefly some statistical properties of the  matrix elements which enter directly the Krylov Hamiltonian Eq.~\ref{eq:H-krylov-tridiag}, namely the  
$\{a_n,b_n\}$ and some correlations therein. As noted in the main text 
(Sec.\ \ref{sec:Model}), it is of course the inevitable existence of correlations in the $\{a_n,b_n\}$ that render the Krylov Hamiltonian fundamentally different from the conventional Anderson model in 1D, despite both being defined on disordered 1D chains with nearest-neighbour hoppings.

We start with the hopping matrix elements, $b_n$. The $n$- and $L$-dependences of the averaged $\braket{b_{n}}$ are shown in the insets to 
Fig.~\ref{fig:krylov-ham}(a1,a2). As demonstrated in the main panels of Fig.~\ref{fig:krylov-ham}(a1,a2), $\braket{b_{n}}$ has the scaling form
\eq{
\braket{b_n^{\pd}} = B(L)\times {\cal B}\left(n/\nh\right)\,
\label{eq:bn-scaling}
}
where $B(L)$ grows with $L$ as $B(L)\sim \sqrt{L}$ and ${\cal B}$ is a decaying function of its argument. 
Moreover, as seen from the figure, this same
scaling behaviour arises for both the weak- and
strong-disorder regimes.
Fluctuations in the $b_{n}$, embodied in $[\braket{b_{n}^{2}}-\braket{b_{n}}^{2}]^{1/2}$, are also found to be relatively small, and not shown here. 

The diagonal matrix elements $\{a_{n}\}$ encode the effective disorder on
the Krylov chain. We find the average $\braket{a_{n}}\simeq 0$ for all 
$n$, so consider the effective disorder as reflected in 
the local standard deviation
\eq{
W_n^{\pd} = \sqrt{\braket{a_n^2} - \braket{a_n}^2}\,.
\label{eq:Wn-def}
}
This too has a a non-trivial scaling with $n$ and $\nh$, but  as shown in 
Fig.~\ref{fig:krylov-ham}(b1,b2) it is seemingly 
somewhat
different in the ergodic and MBL regimes. As seen in Fig.~\ref{fig:krylov-ham}(b1), in the weak disorder ergodic regime $W_n$ is a function solely of $n$.
In the strong disorder regime on the other hand (Fig.~\ref{fig:krylov-ham}(b2)), the diagonal disorder decays on an $n$-scale which seems to scale non-trivially with $\nh$, but 
somewhat
slower than $\nh$.

The on-site disorder on the Krylov chain is of course also correlated. This can be quantified  partially via the covariance
\eq{
 C_{mn}^{\pd} = \braket{a_m^{\pd} a_n^{\pd}}-\braket{a_m^{\pd}}\braket{a_n^{\pd}}\,.
}
However, since the effective disorder at Krylov site $n$ itself depends on $n$, 
it is natural to rescale the correlation as
\eq{
\tilde{C}_{mn}^{\pd} = C_{mn}^{\pd}/(W_{m}^{\pd}W_{n}^{\pd})\,.
\label{eq:an-corr}
}
Numerical results  for $\tilde{C}_{mn}$ are shown in Fig.~\ref{fig:krylov-ham}(c). 
From this it is clear that the correlations are indeed finite, although there is little discernible qualitative difference between the weak- and
strong-disorder regimes.

We do not dwell further on the  scalings of the `bare' matrix elements  $\{a_n,b_n\}$
entering the Krylov Hamiltonian. This is in part because these may depend on the specific Hamiltonian considered. But more importantly, it is because the relative lack of crisp distinctions between them in the weak- and strong-disorder regimes, as noted above, suggests they do not greatly illuminate key diﬀerences between the behaviour of the system in the ergodic and MBL phases.

\bibliography{refs}

\end{document}